\documentclass[amsmath, amssymb, 9.5pt, accepted=2019-03-15]{quantumarticle}

\usepackage{array}

\usepackage{multirow}
\usepackage{listings}
\usepackage{graphicx}
\usepackage{textcomp}
\usepackage{dcolumn}
\usepackage{xcolor}
\usepackage{hyperref}
\hypersetup{
    colorlinks,
    linkcolor={red!50!black},
    citecolor={green!50!black},
    urlcolor={blue!50!black}
}

\newcolumntype{C}[1]{>{\centering\let\newline\\\arraybackslash\hspace{0pt}}m{#1}}
\newcolumntype{L}{>{\centering\arraybackslash}m{3cm}}

\renewcommand{\>}{\rangle}

\usepackage{pifont}
\renewcommand{\check}{\color{green!70!black}\ding{51}}

\definecolor{codegreen}{rgb}{0,0.6,0}
\definecolor{codegray}{rgb}{0.5,0.5,0.5}
\definecolor{codepurple}{rgb}{0.58,0,0.82}
\definecolor{backcolour}{rgb}{0.95,0.95,0.92}
 
\lstdefinestyle{mystyle}{
    backgroundcolor=\color{lightgray!30},   
    commentstyle=\color{codegreen},
    keywordstyle=\color{magenta},
    numberstyle=\tiny\color{codegray},
    stringstyle=\color{codepurple},
    basicstyle=\footnotesize,
    breakatwhitespace=false,         
    breaklines=true,                 
    captionpos=b,                    
    keepspaces=true,                 
    numbers=left,                    
    numbersep=5pt,                  
    showspaces=false,                
    showstringspaces=false,
    showtabs=false,                  
    tabsize=2,
}
 
\begin{document}


\title{Overview and Comparison of Gate Level Quantum Software Platforms}

\author{Ryan LaRose}
\affiliation{Department of Computational Mathematics, Science, and Engineering,
 Michigan State University.}
\affiliation{Department of Physics and Astronomy, Michigan State University}

\date{\today}

\begin{abstract}
Quantum computers are available to use over the cloud, but the recent explosion of quantum software platforms can be overwhelming for those deciding on which to use. In this paper, we provide a current picture of the rapidly evolving quantum computing landscape by comparing four software platforms---Forest (pyQuil), Qiskit, ProjectQ, and the Quantum Developer Kit (Q\#)---that enable researchers to use real and simulated quantum devices. Our analysis covers requirements and installation, language syntax through example programs, library support, and quantum simulator capabilities for each platform. For platforms that have quantum computer support, we compare hardware, quantum assembly languages, and quantum compilers. We conclude by covering features of each and briefly mentioning other quantum computing software packages.
\end{abstract}


\maketitle
\tableofcontents


\section{\label{sec:intro}Introduction}

Quantum programming languages have been thought of at least two decades ago \cite{qprog0,qprog1,qprog2,qprog3,qprog4,qprog5}, but these were largely theoretical and without existing hardware. Quantum computers are now a reality, and there are real quantum programming languages that let anyone with internet access use them. A critical mass of effort from researchers in industry and academia alike has produced small quantum devices that operate on the circuit model of quantum computing. These computers are small, noisy, and not nearly as powerful as current classical computers. But they are nascent, steadily growing, and heralding a future of vast computational power for problems in chemistry \cite{chem0,chem1}, machine learning \cite{ml-ex,ml-ex2}, optimization \cite{opt-example}, finance \cite{finance-ex}, and more \cite{qsim-georgescu}. These devices are a testbed for preparing the next generation of quantum software engineers to tackle current classically intractable computational problems. Indeed, cloud quantum computing has already been used to calculate the deuteron binding energy \cite{deuteron} and test subroutines in machine learning algorithms \cite{state-overlap,qalg-beginners}.

Recently, there has been an explosion of quantum computing software over a wide range of classical computing languages. A list of open-source projects, numbering well over fifty, is available at \cite{mark-github-qsoftware}, and a list of quantum computer simulators is available at \cite{qcsimulators-list}. This sheer number of programs, while positively reflecting the growth of the field, makes it difficult for students and researchers to decide on which software platform to use, getting lost in documentation or being too overwhelmed to know where to start.

In this paper, we hope to provide a succinct overview and comparison of major general-purpose gate-level quantum computing software platforms. From the long list, we have selected four in total: three that provide the user with the ability to connect to real quantum devices--- Forest from Rigetti \cite{pyquil}, Qiskit from IBM \cite{qiskit}, and ProjectQ from ETH Zurich \cite{projectq,projectq2}|and one with similar functionality but no current capability to connect to a quantum computer---the Quantum Development Kit from Microsoft \cite{microsoftqdk}. The ability to connect to a real quantum device has guided our selection of these platforms. Because of this, and for the sake of succinctness, we are intentionally omitting a number of respectable platforms and languages. We briefly mention a few of these in Appendix \ref{app:cirq} and Appendix \ref{sec:other-software}. 

For now, our major goal is to provide a picture of the quantum computing landscape governed by these four platforms. In Section \ref{sec:the-software-platforms}, we cover each platform in turn, discussing requirements and installation, documentation and tutorials, language syntax, and quantum hardware. In Section \ref{sec:comparison}, we provide a detailed comparison of the platforms. This includes quantum algorithm library support in \ref{subsec:library-support}, quantum hardware support in \ref{subsec:hardware-support}, quantum circuit compilers in \ref{subsec:quantum-compilers}, and quantum computer simulators in \ref{subsec:simulator}. We conclude in Section \ref{sec:conclusions} with discussion and some subjective remarks about each platform. Appendix \ref{app:cirq} and Appendix \ref{sec:other-software} discuss other quantum software, Appendix \ref{sec:appendix:simulator-peformance} includes details on testing the quantum circuit simulators, and Appendix \ref{sec:example-programs} shows code for the quantum teleportation circuit in each of the four languages for a side by side comparison. 


\section{\label{sec:the-software-platforms}The Software Platforms}

\begin{figure*}
		\centering
    	\hspace*{-1.7em}
    	\includegraphics[scale=0.57]{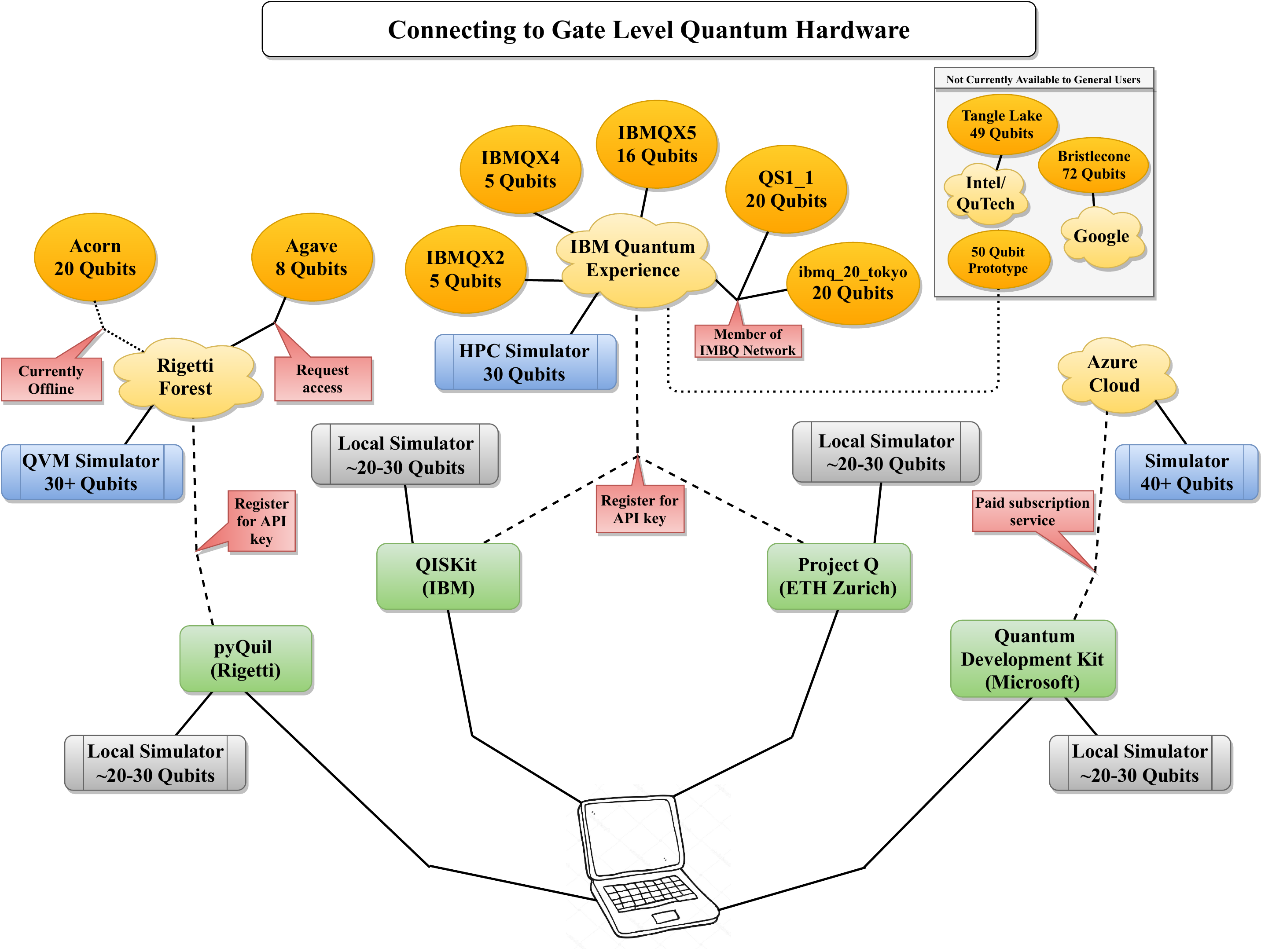}
    	\caption{A schematic diagram showing the paths to connecting a personal computer to a usable gate-level quantum computer. Starting from the personal computer (bottom center), nodes in green shows software platforms that can be installed on the user's personal computer. Grey nodes show simulators run locally (i.e., on the user's computer). Dashed lines show API/cloud connections to company resources shown in yellow clouds. Quantum simulators and usable quantum computers provided by these cloud resources are shown in blue and gold, respectively. Red boxes show requirements along the way. For example, to connect to Rigetti Forest and use the Agave 8 qubit quantum computer, one must download and install pyQuil (available on macOS, Windows, and Linux), register on Rigetti's website to get an API key, then request access to the device. Notes: (i) Rigetti's Quantum Virtual Machine requires an upgrade for more than 30 qubits, (ii) local simulators depend on the user's computer so numbers given are approximates, and (iii) the grey box shows quantum computers that have been announced but are not currently available to general users.}
    	\label{fig:qhardware-diagram}
	\end{figure*}

	An overview of various quantum computers and the software needed to connect to them is shown in Figure~\ref{fig:qhardware-diagram}. 
    At the time of writing, these four software platforms allow one to connect to four different quantum computers---one by Rigetti, an 8 qubit quantum computer which can be connected to via pyQuil; and three by IBM, the largest openly available being 16 qubits, which can be connected to via Qiskit or ProjectQ. There is also a fourth 20 qubit quantum computer by IBM, but this device is only available to members of the IBM Q Network, a collection of companies, universities, and national laboratories interested in and investing in quantum computing%
    \footnote{Members of the IBMQ network include those announced in December 2017---\href{https://www.wired.com/story/why-jp-morgan-daimler-are-testing-computers-that-arent-useful-yet/}{JP Morgan Chase}, \href{https://www.daimler.com/en/}{Daimler}, \href{https://www.samsung.com/us/}{Samsung}, \href{https://www.honda.com/}{Honda}, \href{https://www.ornl.gov/}{Oak Ridge National Lab}, and others---and those announced in April 2018--\href{http://www.zapatacomputing.com/}{Zapata Computing}, \href{https://strangeworks.com/}{Strangeworks}, \href{http://www.qxbranch.com/}{QxBranch}, \href{http://quantumbenchmark.com/}{Quantum Benchmark}, \href{https://qcware.com/}{QC Ware}, \href{https://q-ctrl.com/}{Q-CTRL}, \href{https://cambridgequantum.com/}{Cambridge Quantum Computing (CQC)}, and \href{https://1qbit.com/}{1QBit}. \href{https://www.ncsu.edu/}{North Carolina State University} is the first American university to be a member of the IBM Q Hub, which also includes the \href{http://www.ox.ac.uk/}{University of Oxford} and the \href{https://www.unimelb.edu.au/}{University of Melbourne}. For a complete and updated list, see \href{https://www.research.ibm.com/ibm-q/network/}{https://www.research.ibm.com/ibm-q/network/}.}%
    . Also shown in Figure~\ref{fig:qhardware-diagram} are quantum computers by companies like Google, IBM, and Intel which have been announced but are not currently available. 
    
    The technology of quantum hardware is rapidly changing. It is very likely that new computers will be available by the end of the year, and in two or three years this list may be completely outdated. What will remain, however, is the software used for connecting to this technology. It will be very simple to use these new quantum computers by changing just a few lines of code without changing the actual syntax used for generating or running the quantum circuit. For example, in Qiskit, one could just change the name of the backend when executing the circuit:
    
    	{\footnotesize
    	\begin{lstlisting}[language=Python, caption={The backend specifies which computer (real or simulated) to run quantum programs on using Qiskit. As future quantum computers get released, running on new hardware will be as easy as changing the backend.}]
execute(quantum_circuit, backend=...)\end{lstlisting}
	    }
    \noindent Although the software is changing as well with new version releases%
    \footnote{The programs included in this paper can be found at \href{https://github.com/rmlarose/qsoftware-code}{https://github.com/rmlarose/qsoftware-code} for the most recent version of each platform.}, these are, for the most part, relatively minor syntactical changes that do not alter significantly the software functionality.
    
    At the lowest level of the quantum computing stack, a language must instruct the computer which physical operations to perform on which qubits. We refer to these languages, such as Quil in Forest and OpenQASM in Qiskit, as \textit{quantum assembly/instruction languages}, or occasionally as \textit{quantum languages} for brevity%
    \footnote{To avoid bias towards IBM (which uses the terminology quantum \textit{assembly} language) or Rigetti (which uses the terminology quantum \textit{instruction} language), we abbreviate to \textit{quantum language}.}%
    . On top of quantum languages sit \textit{quantum programming languages}, which are used to manipulate quantum languages in a more natural and readable way for programmers. Examples of quantum programming languages include pyQuil, which is embedded into the classical ``host'' Python programming language, or Q\#, a standalone quantum programming language resembling the classical C\# language. We refer to the collection of a quantum programming language with other tools such as compilers and simulators as a \textit{quantum software platform}, or simply a \textit{platform}. In this paper, we focus on \textit{gate level quantum software platforms} which are designed around the circuit (gate) model of quantum computing.
    
    In what follows, we run through each of the four platforms in turn, discussing requirements and installation, documentation and tutorials, quantum programming language syntax, quantum assembly/instruction language, quantum hardware, and simulator capabilities. Our discussion is not meant to serve as complete instruction in a language, but rather to give the reader a feel of each platform before diving into one (or more) of his/her choosing. Our analysis includes enough information to begin running algorithms on quantum computers. However, we refer the reader, once s/he has decided on a particular platform, to the specific documentation for complete information. We include links to documentation and tutorial sources for each package. We are also assuming basic familiarity with quantum computing, for which many good resources now exist \cite{nielsen-chuang,qcr}.
    
    
    
	\subsection{\label{subsec:pyquil}Forest}
    
    Forest is a quantum software platform developed by Rigetti which includes pyQuil, an open-source quantum programming language embedded in the classical host language Python, for constructing, analyzing, and running quantum programs. pyQuil is built on top of Quil, an open quantum assembly/instruction language designed specifically for near-term quantum computers and based on a shared classical/quantum memory model \cite{forest-rigetti-website} (meaning that both qubits and classical bits are available for memory). Forest also includes the Grove library for algorithms and applications as well as the Reference QVM, a local quantum computer simulator.
    
    \begin{figure}[h!]
		\centering 
        \textbf{Forest Overview} \\
        \begin{tabular}{|C{3.5cm}|C{3.5cm}|} \hline 
        	\textbf{Institution} 			& \href{https://www.rigetti.com/}{Rigetti} \\ \hline 
            \textbf{First Release} 			& v0.0.2 on Jan 15, 2017 \\ \hline 
            \textbf{Version} 		        & v1.9.0 \\ \hline 
            \textbf{Open Source?} 			& \check \\ \hline 
            \textbf{License} 				& Apache-2.0 \\ \hline 
            \textbf{Homepage}			 	& \href{https://www.rigetti.com/index.php/forest}{Home} \\ \hline 
            \textbf{GitHub}					& \href{https://github.com/rigetticomputing/pyquil}{Git} \\ \hline 
            \textbf{Documentation} 			& \href{http://pyquil.readthedocs.io/en/latest/}{Docs}, 
            Tutorials (\href{https://github.com/rigetticomputing/grove}{Grove}) \\ \hline 
            \textbf{OS} 					& Mac, Windows, Linux \\ \hline 
            \textbf{Requirements} 			& \href{https://www.python.org/downloads/}{Python} 3, 
                                    		  \href{https://www.anaconda.com/download/}{Anaconda} (recommended) 	\\ \hline 
            \textbf{Classical Host Language} 	& Python \\ \hline 
            \textbf{Quantum Prog. Lang.} 	& pyQuil \\ \hline 
            \textbf{Quantum Language} 		& \href{http://pyquil.readthedocs.io/en/latest/compiler.html}{Quil} \\ \hline 
            \textbf{Quantum Hardware} 		& 8 qubits \\ \hline 
            \textbf{Simulator} 				& $\sim$20 qubits locally, 26 qubits with most API keys to QVM, 30+ w/ private access \\ \hline 
            \textbf{Features} 				& Generate Quil code, example algorithms in Grove, topology-specific compiler, noise capabilities in simulator, community Slack channel \\ \hline 
        \end{tabular}
	\end{figure}
    
    \paragraph{Requirements and Installation} To install and use pyQuil, Python 3 is required. The Anaconda distribution of Python is recommended for various module dependencies, although it is not required.
    
    The easiest way to install pyQuil is using the Python package manager \textsf{pip}. At a command line on Linux Ubuntu, we type
    
        {\footnotesize
        \begin{lstlisting}[language=bash]
pip install pyquil\end{lstlisting}
        }
    \noindent to successfully install the software. Alternatively, if Anaconda is installed, pyQuil can be installed by typing
    
        {\footnotesize
        \begin{lstlisting}[language=bash]
conda install -c rigetti pyquil\end{lstlisting}
        }
        
    \noindent at a command line. Another alternative is to download the source code from the git repository and install the software this way. To do so, one would type the following commands:
    
        {\footnotesize
        \begin{lstlisting}[language=bash]
git clone https://github.com/rigetti/pyquil
cd pyquil
pip install -e .\end{lstlisting}
        }
    
    \noindent This last method is recommended for any users who may wish to contribute to pyQuil. See the contribution guidelines on Rigetti's GitHub for more information.
    
    \paragraph{Documentation and Tutorials}
    
    Forest has excellent \href{http://pyquil.readthedocs.io/en/latest/index.html}{documentation} hosted online with background information in quantum computing, instructions on installation, basic programs and gate operations, the simulator known as the quantum virtual machine (QVM), the actual quantum computer, and the Quil language and compiler. By downloading the source code of pyQuil from GitHub, one also gets an examples folder with Jupyter notebook tutorials, regular Python tutorials, and a program \textsf{run\_quil.py} which can run text documents written in Quil using the quantum virtual machine. The \href{https://github.com/rigetticomputing/grove}{Grove} library, which can be installed separately from GitHub, contains more examples of quantum algorithms written in pyQuil.
    
    \paragraph{Syntax}
    
    	The syntax of pyQuil is very clean and succinct. The main element for writing quantum circuits is a \textsf{Program} and can be imported from \textsf{pyquil.quil}. Gate operations can be found in \textsf{pyquil.gates}. The \textsf{api} module allows one to run quantum circuits on the virtual machine. One nice feature of pyQuil is that qubit and classical bit registers do not need to be defined a priori but can be rather allocated dynamically. Qubits in the qubit register are referred to by index (0, 1, 2, ...) and similarly for bits in the classical register. A random bit generator circuit, also called a quantum coin flip circuit, can thus be written as follows\footnote{All Forest, Qiskit, and ProjectQ programs in this paper was run and tested on a Dell XPS 13 Developer Edition laptop running 64 bit Ubuntu 16.04 LTS with 8 GB RAM and an Intel Core i7-8550U CPU at 1.80 GHz. A separate computer with a windows environment was used to write and test Q\# programs using Visual Studio Code.}:
        
        {\footnotesize
        \begin{lstlisting}[language=Python, caption={pyQuil code for a random bit generator.}]
# random bit generator circuit in pyQuil
from pyquil.quil import Program
import pyquil.gates as gates
from pyquil import api

qprog = Program()
qprog += [gates.H(0),
           gates.MEASURE(0, 0)]

qvm = api.QVMConnection()
print(qvm.run(qprog), trials=1)\end{lstlisting}
        }
        \noindent In the first three lines, we import the bare minimum needed to declare a quantum circuit/program (line 2), to perform gate operations on qubits (line 3)%
        \footnote{In pyQuil documentation and examples, it is conventional to import only the gates to be used: e.g., \textsf{from pyquil.gates import H, MEASURE}. Here, we import the entire \textsf{pyquil.gates} for comparison to other programming languages, but note that the preferred developer method is the former, which can nominally help speed up code and keep programs cleaner.}, and to execute the circuit (line 4). In line 6 we instantiate a quantum program, and in lines 7-8 we give it a list of instructions: first do the Hadamard gate $H$ to the qubit indexed by 0, then measure the same qubit into a classical bit indexed by 0. In line 10 we establish a connection to the QVM, and in line 11 we run and display the output of the circuit using one ``trial,'' meaning that the circuit is only simulated once. This program prints out, as is standard for pyQuil output, a list of lists of integers zero or one (equivalently, Boolean values): in our case, either \textsf{[[0]]} or \textsf{[[1]]}. In general, the number of elements in the outer list is the number of trials performed. The integers in the inner lists are the final measurements into the classical register. Since we only did one trial, we only get one inner list. Since we only had one bit in the classical register, we only get one integer within this inner list. 
    
    \paragraph{Quantum Language}
    
    The Quil language, analogous to assembly language on classical computers, is what instructs the quantum computer which physical gates to implement on which qubits. The general syntax of Quil is \textsf{GATE index} where \textsf{GATE} is the quantum gate to be applied to the qubit indexed by \textsf{index} (0, 1, 2, ...). pyQuil has a feature for generating Quil code from a given program. For instance, in the above quantum random bit generator, we could add the line
    
    {\footnotesize
        \begin{lstlisting}[language=Python]
print(qprog)\end{lstlisting}
    }
    \noindent at the end to produce the Quil code for the circuit, which is shown below:
    
    {\footnotesize
        \begin{lstlisting}[caption={Quil code for a random bit generator.}]
H 0
MEASURE 0 [0]\end{lstlisting}
    }
        \noindent We note that t is possible to write quantum circuits in a text editor in Quil and then execute the circuit on the QVM using the program \textsf{run\_quil.py}, but writing programs in pyQuil is of course generally easier. One could also modify \textsf{run\_quil.py} to allow circuit execution on the QPU. We remark that the Quil compiler converts a given circuit into Quil code that the actual quantum computer can implement. We will discuss this more in Section \ref{subsec:quantum-compilers}.
    
    \paragraph{Quantum Hardware}
    
    Rigetti has a quantum processor that can be used by those who request access. To request access, one must visit the \href{https://www.rigetti.com/qpu-request}{Rigetti website} and provide a full name, email address, organization name, and description of the reason for QPU access. Once this is done, a company representative will reach out via email to schedule a time to grant the user QPU access%
    \footnote{A new scheme for scheduling time on Rigetti's computers, called Quantum Cloud Services, is in beta testing and may be released in the future as an alternative to the method described in the text.}
    . An advantage of this scheduling process, as opposed to the queue system of Qiskit to be discussed shortly, is that many jobs can be run in the alloted time frame with deterministic runtimes, which is key for variational and hybrid algorithms. These types of algorithms send data back and forth between classical and quantum computers|having to wait in a queue makes this process significantly longer. A (perhaps) disadvantage is that jobs cannot be executed anytime when the QPU is available, but a specific time must be requested and granted.
    
\begin{figure}
    	\centering 
        \includegraphics[scale=0.3]{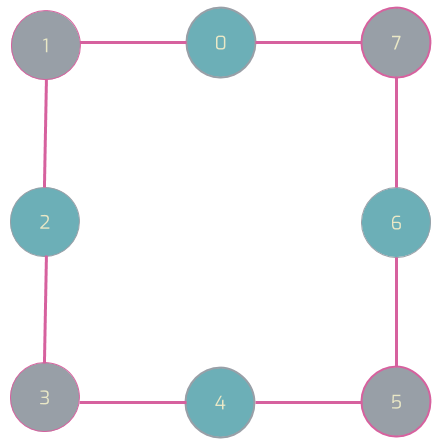}
        \caption{Schematic diagram showing the topology (connectivity) of the 8 qubit Agave QPU by Rigetti. Qubits are labeled with integers 0, 1, ..., 7, and lines connecting qubits indicate that a two qubit gate can be performed between these qubits. For example, we can do Controlled-$Z$ between qubits 0 and 1, but not between 0 and 2. To do the latter, the Quil compiler converts Controlled-$Z$ (0, 2) into operations the QPU can perform. This diagram was taken from pyQuil's documentation.}
        \label{fig:rigetti-acorn}
    \end{figure}

    The actual device, the topology of which is shown in Figure~\ref{fig:rigetti-acorn}, consists of 8 qubits with nearest neighbor connectivity. We will discuss this computer more in detail in Section \ref{subsec:hardware-support}.

    \paragraph{Simulator}
    
    The quantum virtual machine (QVM) is the main utility used to execute quantum circuits. It is a program written to run on a classical CPU that inputs Quil code and simulates the evolution of an actual quantum computer. To connect to the QVM, one must register for an API key for free on \href{https://www.rigetti.com/forest}{https://www.rigetti.com/forest} by providing a name and email address. An email is then sent containing an API key and a user ID which must be set up by running
    
    {\footnotesize
        \begin{lstlisting}[language=Bash]
pyquil-config-setup\end{lstlisting}
        }
    \noindent at the command line (after installing pyQuil, of course). A prompt then appears to enter the emailed keys. 
    
    According to the documentation, most API keys give access to the QVM with up to 30 qubits, and access to more qubits can be requested. The author's API key gives access to 26 qubits (no upgrades were requested). 
    
    Additionally, the Forest library contains a local simulator written in Python and open-sourced, known as the \href{https://github.com/rigetticomputing/reference-qvm}{Reference QVM}. It is not as performant as the QVM, but users can run circuits with as many qubits as they have memory for on their local machines. As a general rule of thumb, circuits with qubits numbering in the low 20s are possible on commodity hardware. The reference QVM must be installed separately, which can be done with \textsf{pip} according to:
    
    {\footnotesize
        \begin{lstlisting}[language=Bash]
pip install referenceqvm\end{lstlisting}
        }
        \noindent To use the Reference QVM instead of the QVM, one simply imports \textsf{api} from \textsf{referenceqvm} instead of from pyQuil:
        
    {\footnotesize
        \begin{lstlisting}[language=Python]
import referenceapi.api as api\end{lstlisting}
        }

	\subsection{\label{subsec:qiskit}Qiskit}
    
    The Quantum Information Software Kit, or Qiskit, is an open-source quantum software platform for working with the quantum language, OpenQASM, of computers in the IBM Q Experience. Qiskit is available in Python, JavaScript, and Swift, but here we only discuss the Python version%
    \footnote{See \href{https://github.com/Qiskit/qiskit-js}{https://github.com/Qiskit/qiskit-js} for information on the JavaScript version and \href{https://github.com/Qiskit/qiskit-swift}{https://github.com/Qiskit/qiskit-swift} for the Swift version.}%
    . Note that the name Qiskit is used interchangeably for the quantum software platform and the quantum programming language.
    
    \begin{figure}[h!]
		\centering 
        \textbf{Qiskit Overview} \\
        \begin{tabular}{|C{3.5cm}|C{3.5cm}|} \hline 
        	\textbf{Institution} 			& IBM \\ \hline 
            \textbf{First Release} 			& 0.1 on March 7, 2017 \\ \hline 
            \textbf{Version} 		        & 0.5.4 \\ \hline 
            \textbf{Open Source?} 			& \check \\ \hline 
            \textbf{License} 				& Apache-2.0 \\ \hline 
            \textbf{Homepage} 				& \href{https://qiskit.org/}{Home} \\ \hline 
            \textbf{Github} 				& \href{https://github.com/Qiskit}{Git} \\ \hline 
            \textbf{Documentation} 			& \href{https://qiskit.org/documentation/}{Docs}, 
            								  \href{https://github.com/Qiskit/qiskit-tutorial}{Tutorial Notebooks}, 		
                                              \href{https://github.com/Qiskit/ibmqx-backend-information/tree/master/backends}{Hardware} \\ \hline 
            \textbf{OS} 					& Mac, Windows, Linux \\ \hline 
            \textbf{Requirements} 			& \href{https://www.python.org/downloads/}{Python 3.5+}, 
            								  \href{https://jupyter.readthedocs.io/en/latest/install.html}{Jupyter Notebooks} (for tutorials),
                                    		  \href{https://www.anaconda.com/download/}{Anaconda 3} (recommended) 	\\ \hline 
            \textbf{Classical Host Language} 	& Python, JavaScript, Swift \\ \hline 
            \textbf{Quantum Prog. Lang.} 	& Qiskit \\ \hline 
            \textbf{Quantum Language} 		& \href{https://github.com/Qiskit/openqasm}{OpenQASM} \\ \hline 
            \textbf{Quantum Hardware} 		& IBMQX2 (5 qubits), IBMQX4 (5 qubits),
            								  IBMQX5 (16 qubits), QS1\_1 (20 qubits) \\ \hline 
            \textbf{Simulator} 				& $\sim$25 qubits locally, 30 through cloud \\ \hline 
            \textbf{Features} 				& Generate QASM code, topology specific compiler, community Slack channel, circuit drawer, Aqua library \\ \hline 
        \end{tabular}
	\end{figure}
    
    \paragraph{Requirements and Installation}
    
    Qiskit is available on macOS, Windows, and Linux. To install Qiskit, Python 3.5+ is required. Additional helpful, but not required, components are Jupyter notebooks for tutorials and the Anaconda 3 Python distribution, which comes with all the necessary dependencies pre-installed.
    
    The easiest way to install Qiskit is by using the Python package manager \textsf{pip}. At a command line, we install the software by typing:
    
    {\footnotesize
        \begin{lstlisting}[language=Bash]
pip install qiskit\end{lstlisting}
    }
        Note that \textsf{pip} automatically handles all dependencies and will always install the latest version. Users who may be interested in contributing to Qiskit can install the source code by entering the following at a command line:
        
        {\footnotesize
        \begin{lstlisting}[language=Bash]
git clone https://github.com/QISKit/qiskit-core
cd qiskit-core
python -m pip install -e .\end{lstlisting}
       }
       \noindent For information on contributing, see the contribution guidelines in Qiskit's online documentation on GitHub.

    \paragraph{Documentation and Tutorials}
    
    The documentation of Qiskit can be found online at \href{https://qiskit.org/documentation/}{https://qiskit.org/documentation/}. This contains instructions on installation and setup, example programs and connecting to real quantum devices, project organization, Qiskit overview, and developer documentation. Background information on quantum computing can also be found for users who are new to the field. A very nice resource is the software development kit (SDK) reference where users can find information on the source code documentation.
    
    Qiskit also contains a large number of tutorial notebooks in a separate GitHub repository (similar to Forest and Grove). These introduce entangled states; standard algorithms like Deutsch-Josza, Grover's algorithm, phase estimation, and the quantum Fourier transform; more advanced algorithms like the variational quantum eigensolver and applications to fermionic Hamiltonians; and even games like ``quantum battleships.'' Additionally, the Aqua library for near-term applications contains example algorithms in fields such as chemistry, finance, and optimization.
    
    There is also very detailed documentation for each of the four quantum backends containing information on connectivity, coherence times, and gate application time. Lastly, we mention the \href{https://quantumexperience.ng.bluemix.net/qx/experience}{IBM Q experience website} and user guides. The website contains a graphical quantum circuit interface where users can drag and drop gates onto the circuit, which is useful for learning about quantum circuits. The user guides contain more instruction on quantum computing and the Qiskit programming language. 
    
    \paragraph{Syntax}
    
    The syntax for Qiskit can be seen in the following example program. In contrast to pyQuil, one has to explicitly allocate quantum and classical registers. Below, we show the program for the random bit circuit:
    
    {\footnotesize
        \begin{lstlisting}[language=Python, caption={Qiskit code for a random bit generator.}]
# random bit generator circuit in Qiskit
from qiskit import QuantumRegister, ClassicalRegister, QuantumCircuit, execute

qreg = QuantumRegister(1)
creg = ClassicalRegister(1)
qcircuit = QuantumCircuit(qreg, creg)

qcircuit.h(qreg[0])
qcircuit.measure(qreg[0], creg[0])

result = execute(qcircuit, backend='local_qasm_simulator', shots=1).result()
print(result.get_counts())\end{lstlisting}
    }
    
    \noindent In line 2 we import the tools to create quantum and classical registers, a quantum circuit, and a function to execute that circuit. We then create a quantum register with one qubit (line 4), classical register with one bit (line 5), and a quantum circuit with both of these registers (line 6). Now that we have allocated quantum and classical registers, we begin providing instructions to construct the circuit: in line 8, we do a Hadamard gate to the zeroth qubit in our quantum register (which is the only qubit in the quantum register); in line 9, we measure this qubit into the classical bit indexed by zero in our classical register (which is the only bit in the classical register)%
\footnote{We could just declare a single qubit and a single classical bit for this program instead of having a register and referring to (qu)bits by index. For larger circuits, it is generally easier to specify registers and refer to (qu)bits by index than having individual names, though, so we stick to this practice here.}.%
\ Now that we have built a quantum circuit, we execute it in line 11 with one ``shot'' (the same as a ``trial'' in pyQuil---the number of times to run the circuit) and print out the result in line 12. By printing \textsf{result.get\_counts()}, we print the ``counts'' of the circuit---that is, a dictionary of outputs and how many times we received each output. For our case, the only possible outputs are 0 or 1, and a sample output of the above program is \textsf{\{'0': 1\}}, indicating that we measured 0 one time (and measured 1 zero times). (Note that the default number of shots in Qiskit is 1024.)

    \paragraph{Quantum Language}
    
    OpenQASM (open quantum assembly language \cite{qasm}) is the quantum language that provides instruction to the actual quantum devices, analogous to Quil and the quantum devices of Forest. The general syntax of OpenQASM is \textsf{gate qubit} where \textsf{gate} specifies a quantum gate operation and \textsf{qubit} labels a qubit. Qiskit has a feature for generating OpenQASM code from a circuit. In the above random bit circuit example, we could add the line
    
    {\footnotesize
        \begin{lstlisting}[language=Python]
print(qcircuit.qasm())\end{lstlisting}
    }
  
    \noindent at the end to produce the QASM code for the circuit, shown below:
    
    {\footnotesize
        \begin{lstlisting}[caption={OpenQASM code for a random bit generator.}]
OPENQASM 2.0;
include "qelib1.inc";
qreg q0[1];
creg c0[1];
h q0[0];
measure q0[0] -> c0[0];\end{lstlisting}
        }
        The first two lines are included in every QASM file. Line 3 (4) creates a quantum (classical) register, and lines 5 and 6 give the instructions for the circuit. It is possible to write small circuits like this directly in OpenQASM, but for larger circuits it is of course easier to have the tools in Qiskit to efficiently program quantum computers. 

    \paragraph{Quantum Hardware}
    
\begin{figure}
    	\centering
        \includegraphics[scale=0.28]{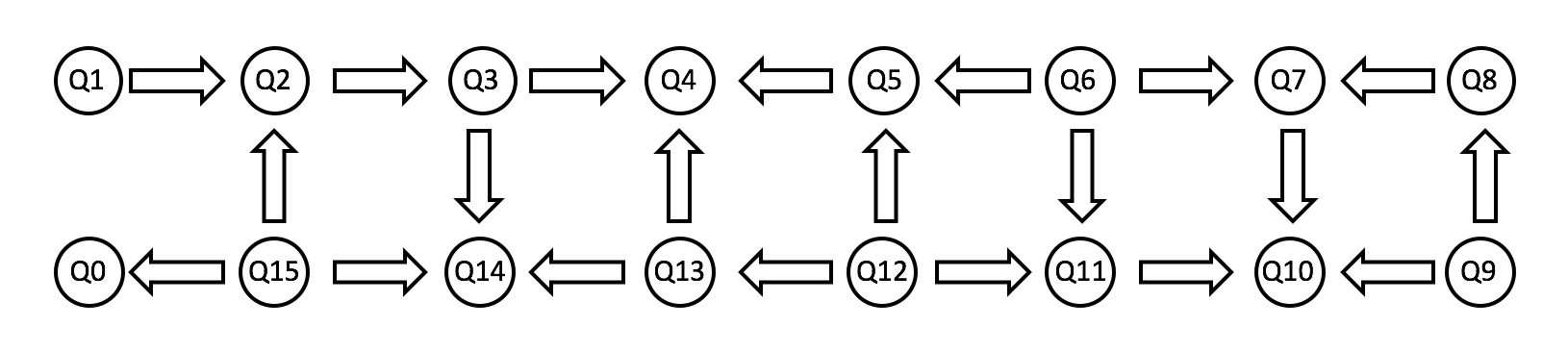}
        \caption{A schematic diagram showing the topology of IBMQX5, taken from \cite{qiskit-ibmqx5-hardware-specs}. Directional arrows show entanglement capabilities. For example, we could perform the operation (in QASM) \textsf{cx Q1, Q2} but not the operation \textsf{cx Q2, Q1}. To do the latter, a compiler translates the instruction into equivalent gates that are performable in the topology and gate set.}
        \label{fig:ibmqx2-connectivity}
    \end{figure}

    There is a vast amount of documentation for the quantum backends supported by Qiskit. These devices include IBMQX2 (5 qubits), IBMQX4 (5 qubits), IBMQX5 (16 qubits), and QS1\_1 (20 qubits, usable only by members of the IBM Q network). Documentation for each is available on GitHub. We discuss in detail IBMQX5 in Section \ref{subsec:hardware-support}, the topology of which is shown in Figure \ref{fig:ibmqx2-connectivity}. 
     
    \paragraph{Simulator} IBM includes several quantum circuit simulators that run locally or on cloud computing resources. These simulators include a local unitary simulator---which applies the entire unitary matrix of the circuit and is limited practically to about 12 qubits---and a state vector simulator---which performs the best locally and can simulate circuits of up to about 25 qubits. For now we just quote qubit numbers, but we discuss the performance of the state vector simulator and compare it to other simulators in Section \ref{subsec:simulator}.
    

	\subsection{ProjectQ}
    
    	ProjectQ is an open-source quantum software platform for quantum computing that features connectivity to IBM's quantum backends, a high performance quantum computer simulator, and several library plug-ins. The first release of ProjectQ was developed by Thomas H\"{a}ner and Damien S. Steiger in the group of Matthias Troyer at ETH Z\"{u}rich, and it has since picked up more contributors. We refer to both the quantum software platform and the quantum programming language as ProjectQ below.

    \begin{figure}[h!]
		\centering 
        \textbf{ProjectQ Overview} \\
        \begin{tabular}{|C{3.5cm}|C{3.5cm}|} \hline 
        	\textbf{Institution} 				& \href{https://www.ethz.ch/en.html}{ETH Zurich} \\ \hline 
            \textbf{First Release} 				& v0.1.0 on Jan 3, 2017 \\ \hline 
            \textbf{Version} 			        & v0.3.6 \\ \hline 
            \textbf{Open Source?} 				& \check \\ \hline 
            \textbf{License} 					& Apache-2.0 \\ \hline 
            \textbf{Homepage} 					& \href{https://projectq.ch/}{Home} \\ \hline 
            \textbf{Github} 					& \href{https://github.com/ProjectQ-Framework/ProjectQ}{Git} \\ \hline 
            \textbf{Documentation} 				& \href{http://projectq.readthedocs.io/en/latest/}{Docs}, 
            									  \href{https://github.com/ProjectQ-Framework/ProjectQ/tree/master/examples}{Example Programs}, 		
                                        		  \href{https://arxiv.org/abs/1612.08091}{Paper} \\ \hline 
            \textbf{OS} 						& Mac, Windows, Linux \\ \hline 
            \textbf{Requirements}			 	& \href{https://www.python.org/downloads/}{Python 2 or 3} \\ \hline 
            \textbf{Classical Host Language} 	& Python \\ \hline
            \textbf{Quantum Prog. Lang.} 	    & ProjectQ \\ \hline 
            \textbf{Quantum Language} 			& --- \\ \hline 
            \textbf{Quantum Hardware} 			& no dedicated hardware, can connect to IBM backends \\ \hline 
            \textbf{Simulator} 					& $\sim$28 qubits locally \\ \hline 
            \textbf{Features} 					& Draw circuits, connect to IBM backends, multiple library plug-ins \\ \hline 
        \end{tabular}
	\end{figure}
    
    \paragraph{Requirements and Installation}
    
    A current version of Python, either 2.7 or 3.4+, is required to install ProjectQ. The documentation contains detailed information on installation for each operating system. In our environment, we do the recommended \textsf{pip} install
    
    {\footnotesize
        \begin{lstlisting}[language=Bash]
python -m pip install --user projectq\end{lstlisting}
        }
    
    \noindent to successfully install the software (as a user). To install via the source code, we can run the following at a command line:
    
    {\footnotesize
        \begin{lstlisting}[language=Bash]
git clone https://github.com/ProjectQ-Framework/ProjectQ
cd projectq
python -m pip install --user .
\end{lstlisting}
    }
    
    \noindent As with previous programs, this method is recommended for users who may want to contribute to the source code. For instructions on doing so, see the contribution guidelines on the ProjectQ GitHub page.
    
    \paragraph{Documentation and Tutorials}
    
    ProjectQ has very good documentation on installation. However, we find the remaining documentation to be a little sparse. The online tutorial provides instruction on basic syntax and example quantum programs (random bits, teleportation, and Shor's factoring algorithm). The rest is the code documentation/reference with information on the structure of the code and each additional module, including functions and classes. The papers \cite{projectq,projectq2} are a good reference and resource, but it is more likely that the online documentation will be more up to date.
    
    \paragraph{Syntax}
    
    
    The syntax of ProjectQ is clear and succinct. The notation for applying gates to qubits is meant to resemble Dirac notation by inserting a vertical line between them. The general construction is \textsf{operation $|$ qubit}, for example $H |0\>$. An example program producing a random bit is shown below.
    
    {\footnotesize
        \begin{lstlisting}[language=Python, caption={ProjectQ code for a random bit generator.}]
# random bit generator circuit in ProjectQ
from projectq import MainEngine
import projectq.ops as ops

eng = MainEngine()
qbits = eng.allocate_qureg(1)

ops.H | qbits[0]
ops.Measure | qbits[0]

eng.flush()
print(int(qbits[0]))\end{lstlisting}
    }
    In line 2, we import the necessary module to make a quantum circuit, and in line 3 we import gate operations. In line 5 we allocate an engine from the \textsf{MainEngine}, and in line 6 we allocate a one qubit register. In lines 8 and 9 we provide the circuit instructions: first do a Hadamard gate on the qubit in the register indexed with a 0, then measure this qubit. This is where the ``quantum syntax'' of the Dirac notation appears within the quantum programming language. We then flush the engine which pushes it to a backend and ensures it gets evaluated/simulated. To mimic the behavior of \textsf{trials} or \textsf{shots} as in pyQuil or Qiskit above, one could wrap lines 6 through 12 in a \textsf{for} loop for the desired number of repetitions. Unlike pyQuil and Qiskit, in ProjectQ one does not specify a classical register when making a measurement. Instead, when we measure \textsf{qbits[0]} in line 9, we get it's value by converting it to an \textsf{int} when we print it out in line 12. (Trying to convert an un-measured qubit to an \textsf{int} throws a \textsf{NotYetMeasuredError} in ProjectQ.) An example output of the program would be printing 0 to the console.
    
    \paragraph{Quantum Language}
    
    As there is no ProjectQ-specific quantum backend, ProjectQ does not have its own dedicated quantum language. If one is using ProjectQ in conjunction with an IBM backend, the code will eventually get converted to OpenQASM, IBM's quantum assembly language.
    
    \paragraph{Quantum Hardware}
    
    ProjectQ does not have its own dedicated quantum computer. One is able to use IBM's quantum backends when using ProjectQ, however.
    
    \paragraph{Simulator}
    
    ProjectQ comes with a fast simulator written in C++, which will be installed by default unless an error occurs, in which case a slower Python simulator will be installed. Additionally, ProjectQ includes a \textsf{ClassicalSimulator} for efficiently simulating stabilizer circuits|i.e., circuits that consist of gates from the normalizer of the Pauli group, which can be generated from Hadmard, CNOT, and phase gates \cite{gottesman-knill}. This simulator is able to handle thousands of qubits to check, e.g., Toffoli adder circuits for specific inputs. However, stabilizer circuits are not universal, so we focus our benchmark and testing on the C++ \textsf{Simulator}.
    
    ProjectQ's C++ \textsf{Simulator} is sophisticated and fast. On the author's computer (the maximum qubit number is limited by the user's local memory, as mentioned), it can handle circuits with 26 qubits of depth 5 in under a minute and circuits of 28 circuits of depth 20 in just under ten minutes. For full details, see section \ref{subsec:simulator} and Figure~\ref{fig:qiskit-projectq-simulator}. 
    
    \paragraph{ProjectQ in other Platforms} ProjectQ is well-tested, robust code and has been used for other platforms mentioned in this paper. Specifically, pyQuil contains ProjectQ code \cite{projectq-in-pyquil}, and the kernels of Microsoft's QDK simulator are developed by Thomas H\"{a}ner and Damian Steiger at ETH Zurich \cite{projectq-in-qdk}, the original authors of ProjectQ. (Note that this does not necessarily mean that the QDK simulator achieves the performance of the ProjectQ C++ simulator as the enveloping code could diminish performance.)
    

	\subsection{\label{subsec:qdk}Quantum Development Kit}
    
    Unlike the superconducting qubit technology of Rigetti and IBM, Microsoft is betting highly on topological qubits based on Majorana fermions. These particles have recently been discovered \cite{majorana} and promise long coherence times and other desirable properties, but no functional quantum computer using topological qubits currently exists. As such, Microsoft currently has no device that users can connect to via their Quantum Development Kit (QDK), the newest of the four quantum software platforms featured in this paper. Nonetheless, the QDK features a new ``quantum-focused'' language called Q\# that has strong integration with Visual Studio and Visual Studio Code and can simulate quantum circuits of up to 30 qubits locally. This pre-release software was first debuted in January of 2018 and, while still in alpha testing, is available on macOS, Windows, and Linux.
    
    \begin{figure}[h!]
		\centering 
        \textbf{QDK Overview}
        \begin{tabular}{|C{3.5cm}|C{3.5cm}|} \hline 
        	\textbf{Institution} 			& Microsoft \\ \hline 
            \textbf{First Release} 			& 0.1.1712.901 on Jan 4, 2018 (pre-release)  \\ \hline 
            \textbf{Version} 		        & 0.2.1802.2202 (pre-release) \\ \hline 
            \textbf{Open Source?} 			& \check \\ \hline 
            \textbf{License} 				& MIT \\ \hline 
            \textbf{Homepage} 				& \href{https://www.microsoft.com/en-us/quantum/development-kit}{Home} \\ \hline 
            \textbf{Github} 				& \href{https://github.com/Microsoft/Quantum}{Git} \\ \hline 
            \textbf{Documentation}			& \href{https://docs.microsoft.com/en-us/quantum/?view=qsharp-preview}{Docs} \\ \hline 
            \textbf{OS} 					& Mac, Windows, Linux \\ \hline 
            \textbf{Requirements} 			& \href{https://code.visualstudio.com/?wt.mc_id=adw-brand&gclid=EAIaIQobChMI3Ijj-KGD2wIVCA1pCh0gQwnIEAAYASAAEgIRNPD_BwE}{Visual Studio Code} (strongly recommended) \\ \hline 
            \textbf{Classical Host Language} 	& C\# \\ \hline
            \textbf{Quantum Prog. Lang.} 	& Q\# \\ \hline 
            \textbf{Quantum Language} 		& --- \\ \hline 
            \textbf{Quantum Hardware} 		& --- \\ \hline 
            \textbf{Simulator} 				& 30 qubits locally, 40 through Azure cloud \\ \hline 
            \textbf{Features} & Built-in algorithms, example algorithms \\ \hline 
        \end{tabular}
	\end{figure}
    
    \paragraph{Requirements and Installation}
    
    	Although it is listed as optional in the documentation, installing Visual Studio Code is strongly recommended for all platforms. (In this paper, we only use VS Code, but Visual Studio is also a possible framework. We remain agnostic as to which is better and use VS Code as a matter of preference.) Once this is done, the version of the QDK can be installed by entering the following at a Bash command line:
        
        {\footnotesize
        \begin{lstlisting}[language=Bash]
dotnet new -i "Microsoft.Quantum.ProjectTemplates::0.2-*"\end{lstlisting}
    	}
    To get QDK samples and libraries from the GitHub repository (strongly recommended for all and especially those who may wish to contribute to the QDK), one can additionally enter:
    
    	{\footnotesize
        \begin{lstlisting}[language=Bash]
git clone https://github.com/Microsoft/Quantum.git
cd Quantum
code .\end{lstlisting}
    	}
    
    \paragraph{Documentation and Tutorials}
    
    The above code samples and libraries are a great way to learn the Q\# language, and the \href{https://docs.microsoft.com/en-us/quantum/?view=qsharp-preview}{online documentation} contains information on validating a successful install, running a first quantum program, the quantum simulator, and the Q\# standard libraries and programming language. This documentation is verbose and contains a large amount of information; the reader can decide whether this is a plus or minus. 
    
    \paragraph{Syntax}
    
    The syntax of Q\# is rather different from the previous three languages. It closely resembles C\# and is more verbose than Python. Shown below is an \textsf{operation}, the analogue of a function in Python, for the same random bit generator circuit that we have shown for all languages. This operation assumes the operation \textsf{Set} is defined, which sets a qubit into a given state.
    
    {\footnotesize
        \begin{lstlisting}[language=C, caption={Q\# code for a random bit generator.}]
// random bit generator circuit in Q#
operation random () : Int
    {
        body
        {
            mutable measured = 0;
            using (qubits = Qubit[1])
            {
                Set (Zero, qubits[0]);
                H(qubits[0]);
                let res = M (qubits[0]);

                // get the measurement outcome
                if (res == One)
                {
                    set measured = 1;
                }
            Set (Zero, qubits[0]);
            }
            // return the measurement outcome
            return measured;
        }
    }\end{lstlisting}
    }
    The use of brackets and keywords can perhaps make this language a little more difficult for new users to learn/read, but at its core the code is doing the same circuit as the previous three examples.
    In line 2 we define an \textsf{operation} (a callable routine with quantum operations) that inputs nothing and returns an integer. Line 4 defines the body of the operation, in which we first initialize the measurement outcome to be zero (line 6) then get a qubit for the circuit (line 7). In line 9, we set the qubit to be the \textsf{Zero} state, perform a Hadamard gate in line 10, then measure the qubit in line 11. Lines 14-17 then grab the measurement outcome which is returned from the operation in line 21. (Line 18 sets the qubit back to the \textsf{Zero} state, which is required in Q\#.) In the Quantum Development Kit, this operation would be saved into a \textsf{.qs} file which contains the Q\# language. A separate \textsf{.cs} ``driver'' file would be used to call the operation, and another \textsf{.csproj} file stores additional meta-data. In total, these three files result in about 60 lines of code. For brevity, we only show the main operation written in Q\# in the \textsf{.qs} file here.\footnote{A complete program with all three files can be found on the GitHub site \href{https://github.com/rmlarose/qsoftware-code}{https://github.com/rmlarose/qsoftware-code} for the most recent version of the QDK.}
    
    Here, we note that the QDK is striving for a high-level language that abstracts from hardware and makes it easy for users to program quantum computers. As an analogy, one does not specifically write out the adder circuit when doing addition on a classical computer---this is done in a high level framework ($a + b$), and the software compiles this down to the hardware level. As the QDK is focused on developing such standards for algorithms involving many gates and qubits, measuring ease of writing code based on simple examples such as a random bit generator and the teleportation circuit (see Appendix \ref{sec:example-programs}) may not do justice to the overall language syntax and platform capabilities, but we include these programs to have some degree of consistency in our analysis.
    
    \paragraph{Quantum Language/Hardware}
    
    As mentioned, the QDK has no current capability to connect to a real quantum computer, and accordingly does not have a quantum assembly/instruction language.
    
    \paragraph{Simulator}
    
    On the user's local computer, the QDK includes a quantum simulator that can run circuits of up to 30 qubits. As mentioned above, kernels for QDK simulators were written by developers of ProjectQ, so performance can be expected to be similar to ProjectQ's simulator performance. (See Section \ref{subsec:simulator}.) Through a paid subscription service to Azure cloud, one can get access to high performance computing that enables simulation of more than 40 qubits. In the QDK documentation, however, there is currently little instruction on how to do this. 
    
    Additionally, the QDK provides a \href{https://docs.microsoft.com/en-us/quantum/quantum-computer-trace-simulator-1?view=qsharp-preview}{trace simulator} that is very effective for debugging classical code that is part of a quantum program as well as estimating the resources required to run a given instance of a quantum program on a quantum computer. The trace simulator allows various performance metrics for quantum algorithms containing thousands of qubits. Circuits of this size are possible because the trace simulator executes a quantum program without actually simulating the state of a quantum computer. A broad spectrum of resource estimation is covered, including counts for Clifford gates, T-gates, arbitrarily-specified quantum operations, etc. It also allows specification of the circuit depth based on specified gate durations. Full details of the trace simulator can be found in the QDK documentation online.
    

	\section{\label{sec:comparison}Comparison}

		Now that the basics of each platform have been covered, in this section we compare each on additional aspects including library support, quantum hardware, and quantum compilers. We also enumerate some notable and useful features of each platform.  
        
        
		\subsection{\label{subsec:library-support}Library Support}
    
    		We use the term ``library support'' to mean examples of quantum algorithms (in tutorial programs or in documentation) or a specific function for a quantum algorithm (e.g., \textsf{language.DoQuantumFourierTransform(...)}). We have already touched on some of these in the previous section. A more detailed table showing library support for the four software platforms is shown in Figure \ref{fig:library-support}. 

			We remark that any algorithm, of course, can be implemented on any of these platforms. Here, we are highlighting existing functionality, which may be beneficial for users who are new to the field or even for experienced users who may not want to program everything themselves. 
            
            As can be seen from the table, pyQuil, Qiskit, and the QDK have a relatively large library support. ProjectQ contains FermiLib, plugins for FermiLib, as well as compatibility with OpenFermion, all of which are open-source projects for quantum simulation algorithms. All examples that work with these frameworks naturally work with ProjectQ. Microsoft's QDK is notable for its number of built-in functions performing these algorithms automatically without the user having to explicitly program the quantum circuit. In particular, the QDK libraries offer detailed iterative phase estimation, an important procedure in many algorithms that can be easily realized on the QDK without sacrificing adaptivity. Qiskit is notable for its large number of tutorial notebooks on a wide range of topics from fundamental quantum algorithms to didactic quantum games.
            
\begin{figure}
	\centering
	{\footnotesize
	\begin{tabular}{|p{2cm}|c|c|c|c|} \hline 
		{\small \textbf{Algorithm}} & \textbf{pyQuil} & \textbf{Qiskit} & \textbf{ProjectQ} & \textbf{QDK} \\ \hline 
		Random Bit Generator			& \check (T) 	& \check (T)	& \check (T)	& \check (T)    \\ \hline 
		Teleportation 					& \check (T) 	& \check (T)	& \check (T)	& \check (T)    \\ \hline 
        Swap Test 						& \check (T) 	& 				& 				& 				\\ \hline 
        Deutsch-Jozsa 			 		& \check (T)    & \check (T)	& 				& \check (T) 	\\ \hline 
		Grover's Algorithm   			& \check (T)	& \check (T)	& \check (T) 	& \check (B)	\\ \hline 
        Quantum Fourier Transform 		& \check (T) 	& \check (T)	& \check (B)	& \check (B)	\\ \hline 
        Shor's Algorithm 				& 				& 				& \check (T) 	& \check (D) 	\\ \hline 
        Bernstein Vazirani	 			& \check (T) 	& \check (T)	& 				& \check (T) 	\\ \hline 
        Phase Estimation	 			& \check (T) 	& \check (T)	& 				& \check (B)	\\ \hline 
        Optimization/ QAOA	 			& \check (T) 	& \check (T)	& 				& 				\\ \hline 
        Simon's Algorithm	 			& \check (T) 	& \check (T)	& 				& 				\\ \hline 
        Variational Quantum Eigensolver	& \check (T) 	& \check (T)	& \check (P)	& 				\\ \hline 
        Amplitude Amplification 		& \check (T) 	& 				& 				& \check (B)	\\ \hline 
        Quantum Walks 					& 				& \check (T)	& 				& 				\\ \hline 
        Ising Solver 					& \check (T) 	& 				& 				& \check (T) 	\\ \hline 
        Quantum Gradient Descent		& \check (T) 	& 				& 				& 				\\ \hline 
        Five Qubit Code 				& 				& 				& 				& \check (B) 	\\ \hline
        Repetition Code 				& 				& \check (T)	& 				& 				\\ \hline
        Steane Code 					& 				& 				& 				& \check (B) 	\\ \hline 
        Draper Adder					& 				& 				& \check (T)	& \check (D)	\\ \hline 
        Beauregard Adder				& 				& 				& \check (T)	& \check (D)	\\ \hline
        Arithmetic 						& 				& 				& \check (B) 	& \check (D)	\\ \hline
        Fermion Transforms 				& \check (T) 	& \check (T) 	& \check (P)	& 				\\ \hline 
        Trotter Simulation				& 				& 				& 				& \check (D)	\\ \hline 
        Electronic Structure (FCI, MP2, HF, etc.) & 	& 				& \check (P)	&  	\\ \hline 
        Process Tomography 				& \check (T) 	& \check (T) 	& 				& \check (D)	\\ \hline 
        Vaidman Detection Test 			& 				& \check (T) 	& 				& 				\\ \hline 
	\end{tabular}
	}
    \caption{A table showing the library support for each of the four software platforms. By ``library support,'' we mean a tutorial notebook or program (T), an example in the documentation (D), a built-in function (B) to the language, or a supported plug-in library (P).}
    \label{fig:library-support}
\end{figure}

    \subsection{\label{subsec:hardware-support}Quantum Hardware}
    
    	In this section we discuss only pyQuil and Qiskit, since these are the only platforms with their own dedicated quantum hardware. Qubit quantity is an important characterization in quantum computers, but equally important---if not more important---is the ``qubit quality.'' By this, we mean coherence times (how ``long qubits live'' before collapsing to bits), gate application times, gate error rates, and the topology/connectivity of the qubits. Ideally, one would have infinite coherence times, zero gate application time, zero error rates, and all-to-all connectivity. In the following paragraphs we document some of the parameters of IBMQX5 and Agave, two of the largest publicly available quantum computers. For full details, please see the online documentation of each platform.
        
        \paragraph{IBMQX5} IBMQX5 is a superconducting qubit quantum computer with nearest neighbor connectivity between its 16 qubits (see Figure~\ref{fig:ibmqx2-connectivity}). The minimum coherence (T2) time is $31 \pm 5$ microseconds on qubit 0 and the maximum is $89 \pm 17$ microseconds on qubit 15. A single qubit gate takes 80 nanoseconds to implement plus a 10 nanosecond buffer after each pulse. \textsf{CNOT} gates take about two to four times as long, ranging from 170 nanoseconds for \textsf{cx q[6], q[7]} to 348 nanoseconds for \textsf{cx q[3], q[14]}. Single qubit gate fidelity is very good at over 99.5\% fidelity for all qubits (fidelity = 1 - error). Multi-qubit fidelity is above 94.9\% for all qubit pairs in the topology. The largest readout error is rather large at about 12.4\% with the average being around 6\%. These statistics were obtained from \cite{qiskit-ibmqx5-hardware-specs}.
        
        Lastly, we mention that to use any available quantum computer by IBM, the user submits his/her job into a queue, which determines when the job gets run. This is in contrast to using Agave by Rigetti, in which users have to request access first via an online form, then schedule a time to get access to the device to run jobs. 
        
        \paragraph{Agave} The Agave quantum computer consists of 8 superconducting transmon qubits with fixed capacitive coupling and connectivity shown in Figure~\ref{fig:rigetti-acorn}. The minimum coherence (T2) time is 9.2 microseconds on qubit 1 and the maximum is 15.52 microseconds on qubit 2. The time to implement a Controlled-$Z$ gate is between 118 and 195 nanoseconds. Single qubit gate fidelity is at an average of 96.2\% (again, fidelity = 1 - error) and minimum of 93.2\%. Multi-qubit gate fidelity is on average 87\% for all qubit-qubit pairs in the topology. Readout errors are unknown. These statistics can be found in the online documentation or through pyQuil. 
    	
    \subsection{\label{subsec:quantum-compilers}Quantum Compilers}
    
      \begin{figure*}
          \centering
          \includegraphics[scale=0.45]{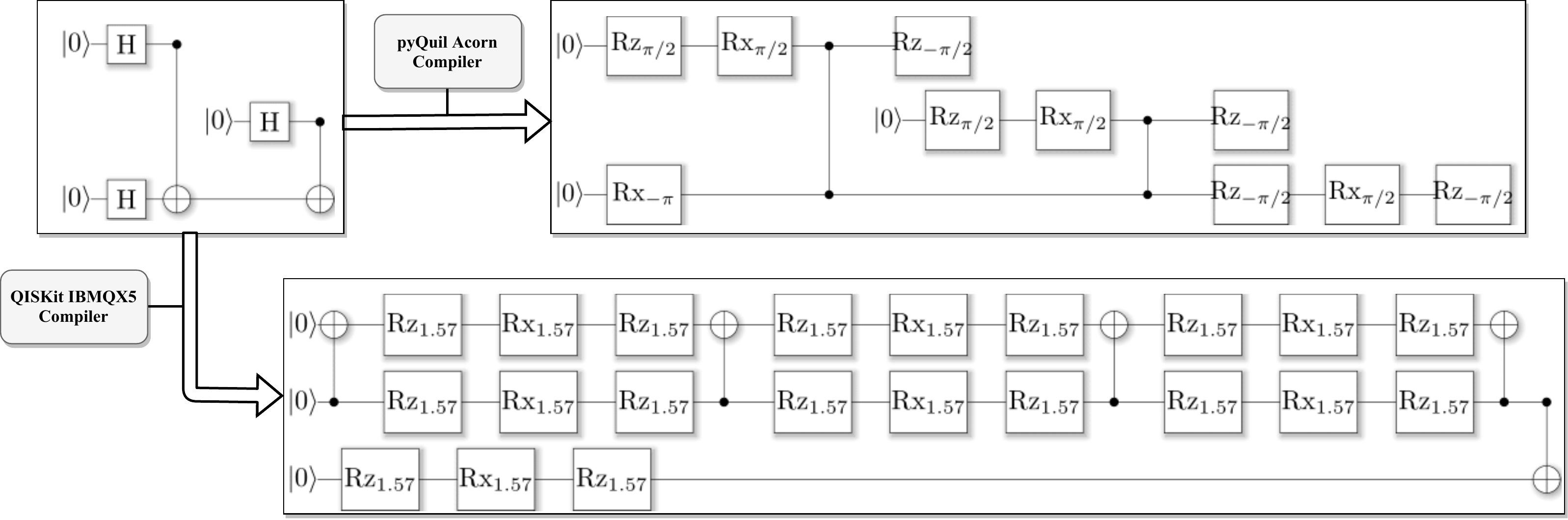}
          \caption{An example of a quantum circuit (top left) compiled by pyQuil for Rigetti's 8 qubit Agave processor (top right), and the same circuit compiled by Qiskit for IBM's 16 qubit IBMQX5. The qubits used on Agave are 0, 1, and 2 (see Figure~\ref{fig:rigetti-acorn}), and the qubits used on IBMQX5 are 0, 1, and 2. Note that neither compiler can directly implement a Hadamard gate $H$ but produces these via products of rotation gates $R_x$ and $R_z$. A \textsf{CNOT} gate can be implemented on IBMQX5, but not on Agave---here, pyQuil must express \textsf{CNOT} in terms of Controlled-$Z$ and rotations. The images of these circuit diagrams were made with ProjectQ.}
          \label{fig:compilers}
      \end{figure*}
    
    	Platforms that provide connectivity to real quantum devices must necessarily have a means of translating a given circuit into operations the computer can understand. This process is known as \textit{compilation}, or more verbosely \textit{quantum circuit compilation}/\textit{quantum compilation}. Each computer has a basis set of gates and a given connectivity|it is the compiler's job to take in a given circuit and return an equivalent circuit obeying the basis set and connectivity requirements. In this section we only discuss Qiskit and Rigetti, for these are the platforms with real quantum computers.
        
        The IBMQX5 basis gates are $u_1$, $u_2$, $u_3$, and \textsf{CNOT} where
        \begin{align*}
        	u_1(\lambda) &= \left[ \begin{matrix}
             	1 & 0 \\
                0 & e^{i \lambda} \\
             \end{matrix} \right] ,\\
            u_2(\phi, \lambda) &= \frac{1}{\sqrt{2}}\left[ \begin{matrix}
            	1 & - e^{i \lambda} \\
                e^{i \phi} & e^{i(\lambda + \phi)} \\
            \end{matrix} \right], \qquad \textrm{and} \\
            u_3(\theta, \phi, \lambda) &= \left[ \begin{matrix}
            	\cos (\theta / 2) & - e^{i \lambda} \sin (\theta / 2) \\
                e^{i \phi} \sin (\theta / 2) & e^{i(\lambda + \phi)} \cos (\theta / 2) \\
            \end{matrix} \right] .
        \end{align*}
         Note that $u_1$ is equivalent to a frame change $R_z (\theta)$ up to a global phase and $u_2$ and $u_3$ are a sequence of frame changes and pulses $R_x (\pi / 2)$
        \begin{align*}
        	u_2 (\phi, \lambda) &= R_z(\phi + \pi / 2) R_x (\pi / 2) R_z(\lambda - \pi / 2), \\
            u_3 (\theta, \phi, \lambda) &= R_z(\phi + 3 \pi) R_x (\pi / 2) R_z( \theta + \pi) R_x (\pi / 2) R_z(\lambda)
        \end{align*}
        with the rotation gates being the standard
        \begin{align*}
        	R_x (\theta) &:= e^{-i \theta X / 2} = \left[ \begin{matrix}
        		\cos \theta / 2 & - i \sin \theta / 2 \\
                - i \sin \theta / 2 & \cos \theta / 2 \\
        	\end{matrix} \right] , \\
            R_z(\theta) &:= e^{- i \theta Z / 2} = \left[ \begin{matrix}
            	e^{- i \theta / 2} & 0 \\
                0 & e^{ i \theta / 2} \\
            \end{matrix} \right]
        \end{align*}
        where $X$ and $Z$ are the usual Pauli matrices. On the IBM quantum computers, $R_z(\theta)$ is a ``virtual gate,'' meaning that nothing is actually done to the qubit physically. Instead, since the qubits are naturally rotating about the $z$-axis, doing a $z$ rotation simply amounts to changing the clock, or frame, of the internal (classical) software keeping track of the qubit. 
        
        The topology of IBMQX5 is shown in Figure~\ref{fig:ibmqx2-connectivity}. This connectivity determines which qubits it is possible to natively perform \textsf{CNOT} gates on, where a matrix representation of \textsf{CNOT} in the computational basis is given by
        \begin{equation*}
        	\textsf{CNOT} := \left[ \begin{matrix}
        		1 & 0 & 0 & 0 \\
                0 & 1 & 0 & 0 \\
                0 & 0 & 0 & 1 \\
                0 & 0 & 1 & 0 \\
        	\end{matrix} \right] .
        \end{equation*}
        Note that it is possible to perform \textsf{CNOT} between any qubits in Qiskit, but when the program is compiled down to the hardware level, the Qiskit compiler converts this into a sequence of \textsf{CNOT} gates allowed in the connectivity. The Qiskit compiler allows one to specify an arbitrary basis gate set and topology, as well as providing a set of parameters such as noise.
        
        For Rigetti's 8 qubit Agave processor, the basis gates are $R_x(k \pi / 2)$ for $k \in \mathbb{Z}$, $R_z(\theta)$, and Controlled-$Z$. The single qubit rotation gates are as above, and the two qubit Controlled-Z (\textsf{CZ}) is given by
        \begin{align*}
            \textsf{CZ} = \left[ \begin{matrix}
        		1 & 0 & 0 & 0 \\
                0 & 1 & 0 & 0 \\ 
                0 & 0 & 1 & 0 \\
                0 & 0 & 0 & -1 \\
        	\end{matrix} \right] .
        \end{align*}
        The topology of Agave is shown in Figure~\ref{fig:rigetti-acorn}. Like Qiskit, pyQuil's compiler also allows one to specify a target instruction set architecture (basis gate set and computer topology). 
        
        An example of the same quantum circuit compiled by both of these platforms is shown in Figure~\ref{fig:compilers}. Here, with pyQuil we compile to the Agave specifications and with Qiskit we compile to the IBMQX5 specifications. As can be seen, Qiskit produces a longer circuit (i.e., has greater depth) than pyQuil. It is not appropriate to claim one compiler is superior because of this example, however. Circuits that are in the language IBMQX5 understands would naturally produce a shorter depth circuit than pyQuil, and vice versa. It is known that any quantum circuit (unitary matrix) can be decomposed into a sequence of one and two qubit gates (see, e.g., \cite{factor-unitary}), but in general this takes exponentially many gates. It is currently a question of significant interest%
        \footnote{IBM's contest ending May 31, 2018, the ``quantum developer challenge,'' is for writing the best compiler code in Python or Cython that inputs a quantum circuit and outputs an optimal circuit for a given topology.}%
         to find an optimal compiler for a given topology.

        \subsection{\label{subsec:simulator}Simulator Performance}
        
        Not all quantum software platforms provide connectivity to real quantum computers, but many platforms include a quantum circuit simulator. This is a program that runs on a classical CPU that mimics (i.e., simulates) the evolution of a quantum computer. As with quantum hardware, it is important to look at not just how many qubits a simulator can handle but also how quickly it can process them, in addition to other parameters like adding noise to emulate quantum computers, etc.
        
        Simulator performance depends on the particular strategy used. Acting on an $n$ qubit state with a $2^n \times 2^n$ matrix requires significantly more memory than just storing the state vector (wavefunction) and acting with one/two qubit gates at a time \cite{state-vector-simulator}. Performance can vary between simulators using the same strategy due to minor differences in program execution, what underlying libraries are used to perform matrix algebra, whether multi-threading is used, etc. In this section, we evaluate the performance of Qiskit's local state vector simulator and ProjectQ's local C++ simulator using the program listed in Appendix \ref{sec:appendix:simulator-peformance}. Both of these programs use the general strategy of only storing the state vector of the system. First, we mention the performance of pyQuil's QVM simulator. 
        
        \paragraph{pyQuil} The Rigetti simulator, called the Quantum Virtual Machine (QVM), does not run on the users local computer but rather through computing resources in the cloud. As mentioned, this requires an API key to connect to. Most API keys give access to 30 qubits initially, and more can be requested. The author is able to simulate a 16 qubit circuit of depth 10 in 2.61 seconds on average. A circuit size of 23 qubits of depth 10 was simulated in 56.33 seconds, but no larger circuits could be simulated because the QVM terminates after one minute of processing with the author's current API access key. 
        
        The QVM contains sophisticated and flexible noise models to emulate the evolution of an actual quantum computer. This is key for developing short depth algorithms on near term quantum computers, as well as for predicting the output of a particular quantum chip. Users can define arbitrary noise models to test programs, in particular define noisy gates, add decoherence noise, and model readout noise. For full details and helpful example programs, see the \href{http://pyquil.readthedocs.io/en/latest/noise.html}{Noise and Quantum Computation} section of pyQuil's documentation.
        
        \paragraph{Qiskit} Qiskit has several quantum simulators available as backends: the \textsf{local\_qasm\_simulator}, the \textsf{local\_state\_vector\_simulator}, the \textsf{ibmq\_qasm\_simulator}, the \textsf{local\_unitary\_simulator}, and the \textsf{local\_clifford\_simulator}. The differences in these simulators is the strategy of simulating quantum circuits. The unitary simulator implements basic (unitary) matrix multiplication and is limited quickly by memory. The state vector simulator does not store the full unitary matrix but only the state vector and single/multi qubit gate to apply. Both methods are discussed in \cite{state-vector-simulator}, and \cite{45qubit-simulation,yaoyun-simulator,adv-sim-compression-overview-2017} contains details on other techniques. Similar to the discussion of the \textsf{ClassicalSimulator} in ProjectQ, the \textsf{local\_clifford\_simulator} is able to efficiently simulate stabilizer circuits, which are not universal. 
        
        Using the local unitary simulator, a circuit of 10 qubits on depth 10 is simulated in 23.55 seconds. Adding one more qubit increases this time by approximately a factor of ten to 239.97 seconds, and at 12 qubits the simulator timed out after 1000 seconds (about 17 minutes). This simulator quickly reaches long simulation times and memory limitations because for $n$ qubits, the unitary matrix of size $2^{n} \times 2^{n}$ has to be stored in memory.
        
\begin{figure}
    	\centering
        \hspace*{0.50em} \textbf{Qiskit State Vector Simulator Performance} \\
        \includegraphics[scale=0.22]{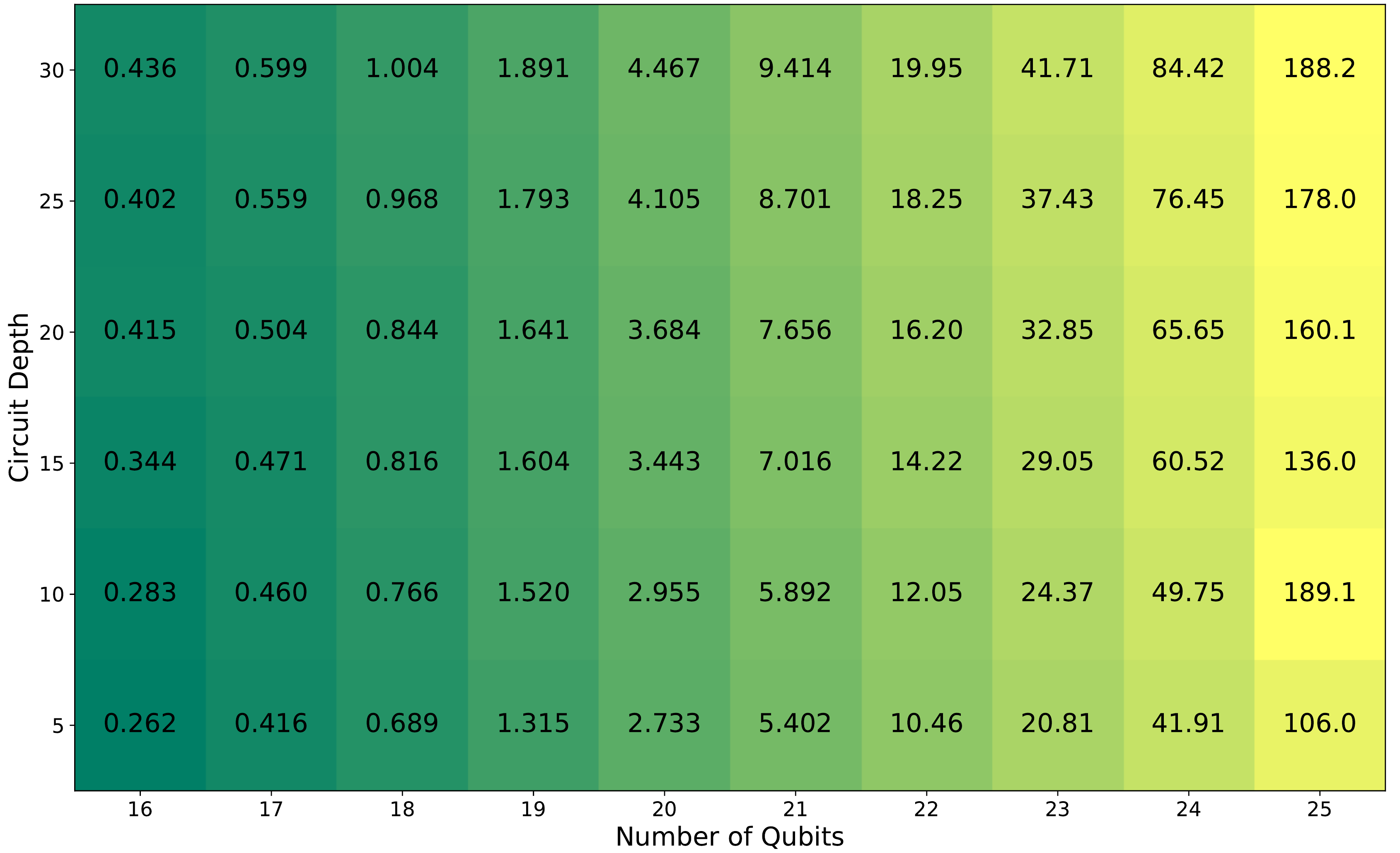}
        \hspace*{0.80em} \textbf{ProjectQ C++ Simulator Performance} \\
        \includegraphics[scale=0.20]{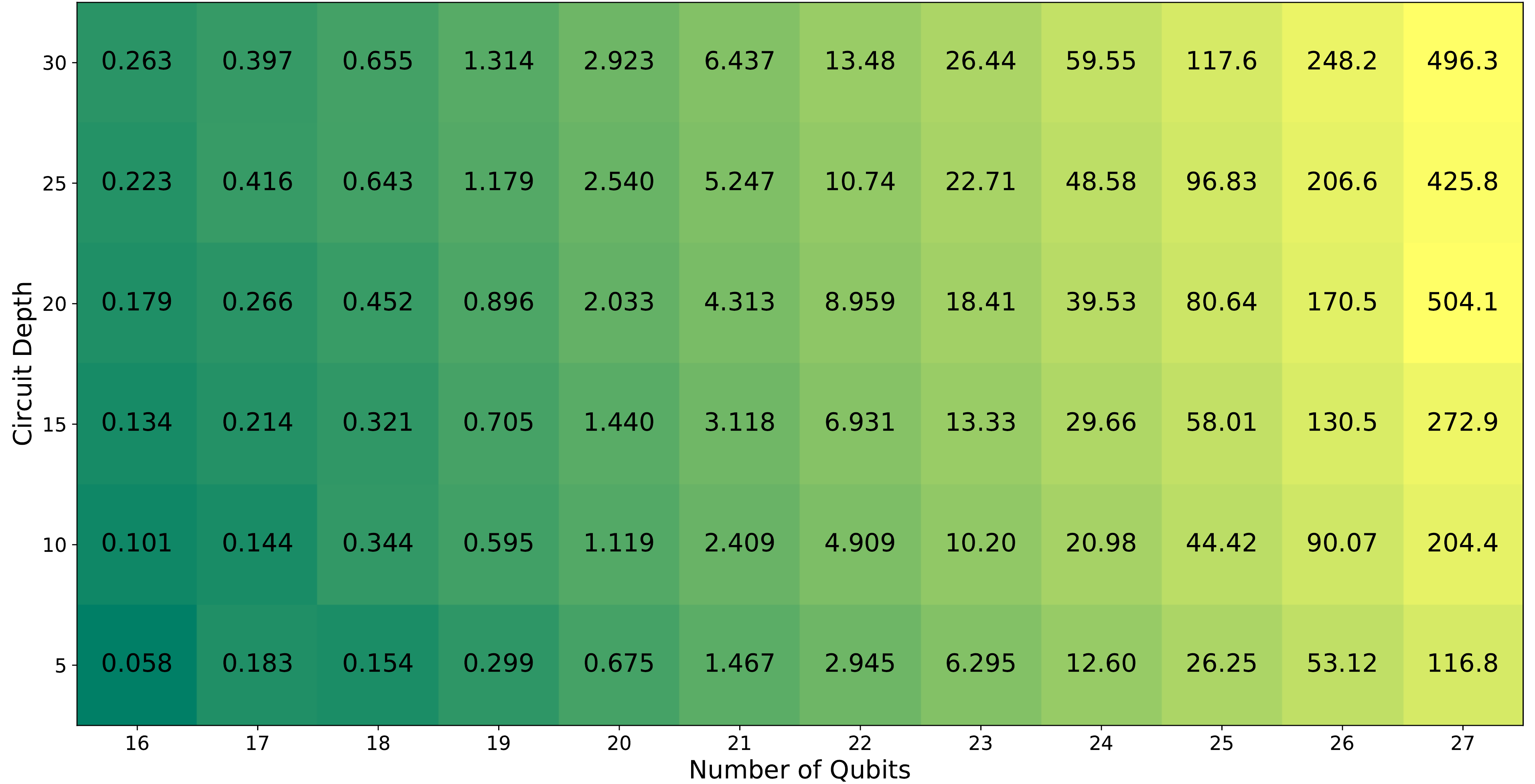}
        \caption{Plots of the performance of Qiskit's local state vector simulator (top) and  ProjectQ's C++ simulator (bottom), showing runtime in seconds for a given number of qubits (horizontal axis) and circuit depth (vertical axis). Darker green shows shorter times and brighter yellow shows longer times (color scales are not the same for both plots). For more details on the testing, see Appendix \ref{sec:appendix:simulator-peformance}.}
        \label{fig:qiskit-projectq-simulator}
    \end{figure}

\begin{figure}
        \centering 
        \includegraphics[scale=0.4]{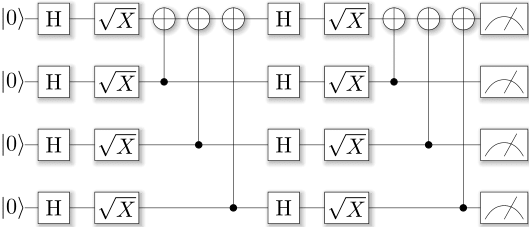}
        \caption{The circuit used for testing the ProjectQ C++ simulator and Qiskit local state vector simulator, shown here on four qubits. In the actual testing, the pattern of Hadamard gates, $\sqrt{X}$ gates, then the sequence of \textsc{CNOT} gates defines one level in the circuit. This pattern is repeated until the desired depth is reached. This image was produced using ProjectQ.}
        \label{fig:projectq-simulator-test-circuit}
	\end{figure}

        The state vector simulator significantly outperforms the unitary simulator. We are able to simulate circuits of 25 qubits in just over three minutes. Circuits of up to 20 qubits with depth up to thirty are all simulated in under five seconds. See Figures~\ref{fig:qiskit-projectq-simulator} and \ref{fig:projectq-simulator-test-circuit} for complete details.  
        
        \paragraph{ProjectQ} ProjectQ comes with a high performance C++ simulator that performed the best in our local testing. The maximum size circuit we were able to successfully simulate was 28 qubits, which took just under ten minutes (569.71 seconds) with a circuit of depth 20. For implementation details, see \cite{projectq}. For the complete performance and testing, see Figures~\ref{fig:qiskit-projectq-simulator} and \ref{fig:projectq-simulator-test-circuit}.
        
        \paragraph{QDK} Although we do not test the QDK simulators here, we note that performance can be expected to be similar to performance of ProjectQ's simulators as the underlying kernels for the QDK simulator were developed by the ProjectQ developers \cite{projectq-in-qdk}.

        \subsection{\label{subsec:features}Features}
        
        A nice feature of Forest is Grove, a separate GitHub repository that can be installed containing tutorials and example algorithms using pyQuil. Rigetti is also building a solid community of users as exemplified by their dedicated Slack channel for Rigetti Forest. The Quil compiler and it's ability to compile for any given instruction set architecture (topology and gate basis) are also nice features. Lastly, pyQuil is compatible with OpenFermion \cite{openfermion}, an open-source Python package for compiling and analyzing quantum algorithms to simulate fermionic systems, including quantum chemistry.
        
        Qiskit is also available in JavaScript and Swift for users who may have experience in these languages. For beginners, Python is a very good starter programming language because of its easy and intuitive syntax. Like Grove, Qiskit also contains a dedicated repository of example algorithms and tutorials. Additionally, the Aqua library in Qiskit contains numerous algorithms for quantum chemistry and artificial intelligence. This library can be run through a graphical user interface or from a command line interface. IBM is second to none for building an active community of students and researchers using their platform. The company boasts of over 3 million remote executions on cloud quantum computing resources using Qiskit run by more than 80,000 registered users, and there have been more than 60 research publications written using the technology \cite{qiskit-published-research}. Qiskit also has a dedicated Slack channel with the ability to see jobs in the queue, a useful feature for determining how long a job submission will take to run. Additionally, the newest release of Qiskit contains a built-in circuit drawer.
        
        Likewise, ProjectQ contains a circuit drawer. By adding just a few lines of code to programs, one can generate TikZ code to produce high quality \TeX \ images. All quantum circuit diagrams in the main text of this paper were made using ProjectQ. The local simulator of ProjectQ is also a great feature as it has very high performance capabilities. Although ProjectQ has no dedicated quantum hardware of its own, users are able to connect to IBM's quantum hardware. Additionally, ProjectQ has multiple library plug-ins including OpenFermion, as mentioned above.   
        
        The QDK was available exclusively on Windows until it received support on macOS and Linux in February 2018. The capability to implement quantum algorithms without explicitly programming the circuit is a nice feature of the QDK, and there are also many good tutorials in the documentation and examples folder for quantum algorithms. It is also notable that Q\# provides auto-generation features for, e.g., the adjoint or controlled version of a quantum operation. In a more general sense, the QDK emphasizes and offers important tools for productive quantum algorithm development including the testing of quantum programs, estimating resource requirements, programming on different models of quantum computation targeted by different hardware, and ensuring the correctness of quantum programs at compile time. These aspects are key in moving towards high-level quantum programming languages.
        

\section{\label{sec:conclusions}Discussion and Conclusions}

	At this point, we hope that the reader has enough information and understanding to make an informed decision of what quantum software platform(s) is (are) right for him/her. A next step is to begin reading the documentation of a platform, install it, and begin coding. In a short amount of time one can begin running algorithms on real quantum devices and begin researching/developing applications in their respective field. 
    
    For those who may be still undecided, we offer the following subjective suggestions. As Python is generally an easier language to pick up than C-style languages, either Forest, Qiskit, or ProjectQ may be more appropriate for beginners. For those with experience in C\#, the Quantum Development Kit may be easier to pick up. Forest, Qiskit, and the QDK all contain good resources for learning about quantum computing. To test algorithms on real quantum computers, Forest and Qiskit are the obvious choices. ProjectQ is great for simulating algorithms on a large number of qubits.
        
        
        
        
        
    
    Again, these are simply suggestions and we encourage the reader to make his/her own choice. All platforms are significant achievements in the field of quantum computing and excellent utilities for students and researchers to program real quantum computers. As a final remark, we note that there are additional quantum software packages being developed, a few of which are mentioned in Appendix \ref{app:cirq} and Appendix \ref{sec:other-software}. 


\section{\label{sec:acknowledgements}Acknowledgements}

We acknowledge use of the IBM Q experience for this work. The views expressed are those of the authors and do not reflect the official policy or position of IBM or the IBM Q experience team. RL thanks Doug Finke for many useful comments and edits on earlier drafts of this paper. We thank developers for feedback and helping us achieve a more accurate representation of each respective platform. Specifically, we thank Damian Steiger from ETH Z\"{u}rich; Will Zeng from Rigetti; and Cathy Palmer, Julie Love, and the Microsoft Quantum Team from Microsoft. RL acknowledges support from Michigan State University through an Engineering Distinguished Fellowship.



\appendix


\section{\label{app:cirq}Cirq}

    	    Cirq is a platform for working with quantum circuits on near-term quantum computers that is very similar in purpose to the four platforms reviewed in the main text%
	    \footnote{The reason Cirq is not included in the main text is because it was released after the first version of this paper. For an article in which Cirq code can be run interactively, see \href{https://github.com/rmlarose/cirq-overview}{https://github.com/rmlarose/cirq-overview}.}%
	    .In this section we analyze Cirq in a similar fashion to these platforms. It should be noted that Cirq is still in alpha testing and will likely release breaking changes in future releases. The following description is for version 0.4.0.
	    
	    Cirq is available on all three major operating systems and requires a working installation of Python 3. The easiest way to install Cirq is by typing
	   
        \begin{lstlisting}[language=bash]
pip install cirq\end{lstlisting}
        
        \noindent at a command line. Alternatively, the source code can be downloaded from \cite{cirq}. This method is recommended for users who may wish to contribute to the platform, for which detailed contribution guidelines exist.
        
        The documentation for Cirq is noticeably more sparse than other platforms, especially in terms of tutorials and examples. Currently, there exists a detailed tutorial on implementing the variational quantum eigensolver using Cirq and a few other tutorial scripts on selected topics. The documentation also contains information on using circuits, gates, and quantum computer simulators.
        
        To get an idea for the language syntax, we include the same random bit generator program in Cirq below:
        




        \begin{lstlisting}[language=Python, caption={Cirq code for a random bit generator.}]
import cirq

qbits = [cirq.LineQubit(0)]
circ = cirq.Circuit()

circ.append(cirq.H(qbits[0]))
circ.append(cirq.measure(qbits[0], key="z")

simulator = cirq.Simulator()
result = simulator.run(circ, repetitions=1)\end{lstlisting}
    
    In this short program, we import the Cirq library in line 2, then create a qubit register and circuit in lines 4-5. Qubit registers are stored simply as lists (more generally, iterables) of qubits. Since Cirq is focused on near-term quantum computing, qubits come as \textsf{LineQubit}s or \textsf{GridQubit}s, as these are common constructions in near-term architectures. In line 7 we apply a Hadamard gate to the qubit, and in line 8 we measure the qubit in the computational basis. (The \textsf{key} is an optional argument that is useful for obtaining measurement results.) Finally, in line 10 we get a quantum computer simulator and use it to run the circuit in line 12 for a total of one \textsf{repetition} (the same as \textsf{trials} or \textsf{shots} in pyQuil or Qiskit). The outcome of the circuit could then be obtained from \textsf{result.histogram(key="z")}, which would return a \textsf{Counter} object of key-value pairs corresponding to measurement outcome and frequency of occurrence. An example output could thus be \textsf{Counter({1: 1})}, indicating that the bit 1 was measured one time.
    
    As expressed in the documentation, Cirq will be an interface for researchers to use their 22 qubit Foxtail and 72 qubit Bristlecone quantum computers. However, these are not yet available to general users over the platform. As such, Cirq does not currently have its own quantum language for communicating with quantum processors. (There is functionality to output OpenQASM code for running on IBM's quantum computers, however.)
    
    As shown in the random bit generator program above, Cirq does include quantum computer simulators. The \textsf{Simulator} used above works for generic gates that implement their unitary matrix, and there is also an \textsf{XmonSimulator} that is specialized to the native gate set of Google's quantum computers. Neither of these simulators contain noise capabilities, but both are able to emulate the (noiseless) behavior of running on a quantum computer or access the wavefunction for debugging purposes.
    
    Notable features of Cirq include built-in utilities for optimizing quantum circuits by reducing the number of gates, automatic hardware-specific compilation, and several useful tools for working with variational hybrid algorithms such as parameterized gates and the ability to simulate a ``sweep'' of the parameters, i.e. a particular set of angles in parameterized gates. This simplifies and can speed up the optimization process because a new quantum circuit does not have to be created after every optimization pass.
    
    Other features include a text-based circuit drawer useful for debugging and the ability to print out quantum states in Dirac notation. Moreover, Cirq allows programmers to define \textsf{Schedules} and \textsf{Devices} to work at the lowest level of algorithm execution, for example specifying the duration of pulses and gates. The ability to simulate noisy quantum circuits is being developed in Cirq and will likely be released in future versions.
    
    For comparison with other languages in Appendix \ref{sec:example-programs}, we include a complete program in Cirq for the quantum teleportation algorithm here. To the author's knowledge, there does not exist an easy way to perform classical conditional operations in Cirq such as those required by the teleportation algorithm. An alternative approach using Cirq's ability to compute reduced density matrices is used in the program below.
    
    \begin{lstlisting}[language=Python]
#!/usr/bin/env python3
# -*- coding: utf-8 -*-

# ========================================
# teleport.py
#
# Teleportation circuit in Cirq.
# ========================================

# ----------------------------------------
# imports
# ----------------------------------------

import cirq

# ----------------------------------------
# qubits and circuit
# ----------------------------------------

qbits = [cirq.LineQubit(x) for x in range(3)]
circ = cirq.Circuit()

# ----------------------------------------
# teleportation circuit
# ----------------------------------------

# perform X to teleport |1> to qubit three
circ.append(cirq.ops.X(qbits[0]))

# main circuit
circ.append([cirq.ops.H(qbits[1]),
    cirq.ops.CNOT(qbits[1], qbits[2]),
    cirq.ops.H(qbits[0]),
    cirq.ops.CNOT(qbits[0], qbits[1]),
    cirq.measure(qbits[0]),
    cirq.measure(qbits[1])])

# print the circuit
print(circ)

# --------------------------------------
# compute the reduced state of qubit three
# --------------------------------------

# get a simulator
simulator = cirq.google.XmonSimulator()

# simulate the circuit with access to the wavefunction
res = simulator.simulate(circ)

# print out the density matrix, which should be
# [[0, 0],
#  [0, 1]]
print(res.density_matrix([2]))\end{lstlisting}

\section{\label{sec:other-software}Other Quantum Software}

	As mentioned in the main text, it would be counterproductive to include an analysis of all quantum software platforms or quantum computing companies. For an updated and current list, see \cite{mark-github-qsoftware}. Our selections in this paper were largely guided by the ability for general users to connect to and use real quantum devices, as well as unavoidable factors like the author's experience and release date of the software platform. In this appendix, we briefly mention other software platforms. The first three are predecessors to the quantum software platforms presented in the main text, and the remaining are other modern quantum software platforms still being developed.
	
	\paragraph{Quipper}
	
	    Quipper is a functional quantum programming language developed by Peter Selinger, Richard Eisenberg, et al. \cite{quipper-website, quipper-paper}. Like pyQuil, Qiskit, and ProjectQ are embedded into the classical Python programming language, Quipper is embedded into Haskell, a statically typed and purely functional classical language.
	    
	    The Quipper website \cite{quipper-website} contains detailed documentation on installation and the language itself. For an example of syntax, the following code snippet, taken from \cite{quipper-intro}, writes a function that inputs a Boolean value (0 or 1), creates a qubit corresponding to this value ($|0\>$ or $|1\>$), and acts on the qubit with the Hadamard gate, returning the result:
	    
	    {\footnotesize \begin{lstlisting}[language=Haskell]
plus_minus :: Bool -> Circ Qubit
plus_minus b = do
q <- qinit b
r <- hadamard q
return r\end{lstlisting}}
	    
	    A complete description of this program, as well as many other example programs, can be found in \cite{quipper-intro}. Quipper has many built-in libraries for quantum computing subroutines and algorithms, for example quantum linear systems, finding unique shortest vectors, and ground state estimation. The \textsf{QuipperLib} consists of additional modules that can be used but is not part of the Quipper programming language proper. (In this sense, Quipper and the QuipperLib can be thought of as a quantum software platform.) Some examples of libraries include \textsf{Qram} for efficient implementation of random access memory and \textsf{QFT} which contains an implementation of the Quantum Fourier Transform. See the documentation for full details and library support as well as additional features such as abilities to draw quantum circuits and automatically generate reversible circuits from ordinary functional programs.
	    
	    Finally, we mention that Quipper does not currently provide any support for connecting to quantum computers, though ``it was designed to control an actual (future) quantum computer'' \cite{quipper-intro}. Quipper does have the ability to simulate classical circuits, stabilizers circuits, and quantum circuits, however. At the time of writing, the latest release of Quipper (v0.8) was in 2016 \cite{quipper-website}.
	    
	\paragraph{Scaffold} 
	
	    Scaffold is a quantum programming language embedded into the classical language C \cite{scaffold-paper}. It is a pure quantum programming language in that its main purpose is to assist in writing quantum algorithms, not necessarily running or simulating them. A typical Scaffold program has elements familiar to C/C++ programmers---preprocessor directives and a \textsf{main} module (function)---as well as elements familiar to quantum computer scientists---quantum gates and qubit registers. An example of a simple short quantum program creating a quantum register and applying a Hadamard gate is shown below.
	    
	    {\footnotesize \begin{lstlisting}[language=C]
#include "gates.h"
module main () {
    int i=0;
    qreg qubit[1];
    H(qubit[i]);
}\end{lstlisting}}
	    
	    The standard library \textsf{gates.h} includes definitions of commonly used gates in quantum computing such as the Hadamard used above. The Scaffold language allows for both ``classical data types'' (e.g. arrays, structs, and unions) and ``quantum data types'' (e.g.,  qubit registers, quantum structs, and quantum unions). The \textsf{Classical to Quantum Gates} module type (\textsf{c2qg}) allows programmers to give classical descriptions of quantum instructions at a higher level, making it easier to program quantum circuits. For example, one can write a module for the Toffoli gate in terms of its effect on the target qubit conditioned on the control qubits, rather than 15 Hadamard, CNOT, and $T$ gates on these qubits.
	    
	    Scaffold sits at the highest level of the quantum computing stack and serves as an interface between quantum programmers and quantum compilers. The ScaffCC library \cite{scaffold-compiler} is an open-source compiler and scheduler written for Scaffold, meant to input quantum algorithms written in Scaffold and output a compiled algorithm, among other utilities. Scaffold programs can be compiled to OpenQASM and the QX quantum computer simulator\footnote{See \href{https://qutech.nl/qx-quantum-computer-simulator/}{https://qutech.nl/qx-quantum-computer-simulator/}.}. At the time of writing, the latest release of StaffCC is version 4.0.
	
	\paragraph{QCL} The QCL (Quantum Computing Language) \cite{qcl-website} is the original quantum programming language and first, to the author's knowledge, of its kind. The QCL, a C-style language last updated in 2014, contains many data types and other constructs that modern quantum programming languages inherit, such as qubits, quantum registers, sub-registers, quantum operations, and so on. The language contains both classical and quantum control flow, ``pseudo-classical'' operators, and abilities to implement query transformations required for ``black box'' algorithms like the standard Deutsch-Jozsa and Bernstein-Vazirani algorithms. Short example programs to get an idea for the QCL syntax, as well as longer programs implementing algorithms like Grover and Shor, can be found in \cite{qcl-paper}.

	\paragraph{Strawberry Fields}
    
    	Developed by the Toronto-based startup \href{https://www.xanadu.ai/}{Xanadu}, Strawberry Fields is a full-stack quantum software platform for designing, optimizing, and simulating quantum optical circuits \cite{strawberry-fields}. Xanadu is developing photonic quantum computers with continuous variable qubits, or ``qumodes'' (as opposed to the discrete variable qubits), and though the company has not yet announced an available quantum chip for general users, one may be available in the near future. Strawberry Fields has a built in simulator using Numpy and TensorFlow, and a quantum programming language called Blackbird. One can download the source code from GitHub, and example tutorials can be found for quantum teleportation, boson sampling, and machine learning. Additionally, the Xanadu website \href{https://www.xanadu.ai/}{https://www.xanadu.ai/} contains an interactive quantum circuit where users can drag and drop gates or choose from a library of sample algorithms.
    
    

\section{\label{sec:appendix:simulator-peformance} Testing Simulator Performance}

Below is the listing of the program for testing the ProjectQ C++ local simulator performance. These tests were performed on a Dell XPS 13 Developer Edition running 64 bit Ubuntu 16.04 LTS with 8 GB RAM and an Intel Core i7-8550U CPU at 1.80 GHz.

{\footnotesize
\begin{lstlisting}[language=Python]
# ----------------------------------------
# imports
# ----------------------------------------

from projectq import MainEngine
import projectq.ops as ops
from projectq.backends import Simulator
import sys
import time

# ----------------------------------------
# number of qubits and depth
# ----------------------------------------

if len(sys.argv) > 1:
    n = int(sys.argv[1])
else:
    n = 16

if len(sys.argv) > 1:
    depth = int(sys.argv[2])
else:
    depth = 10 

# ----------------------------------------
# engine and qubit register
# ----------------------------------------

eng = MainEngine(backend=Simulator(gate_fusion=True), engine_list=[])
qbits = eng.allocate_qureg(n)

# ----------------------------------------
# circuit
# ----------------------------------------

# timing -- get the start time
start = time.time()

# random circuit
for level in range(depth):
    for q in qbits:
        ops.H | q
        ops.SqrtX | q
        if q != qbits[0]:
            ops.CNOT | (q, qbits[0])

# measure
for q in qbits:
    ops.Measure | q

# flush the engine
eng.flush()

# timing -- get the end time
runtime = time.time() - start

# print out the runtime
print(n, depth, runtime)\end{lstlisting}
}

The circuit, which was randomly selected, is shown in Figure~\ref{fig:projectq-simulator-test-circuit}. We remark that the Qiskit simulator was tested on an identical circuit---we omit the code for brevity.


	\section{\label{sec:example-programs}Example Programs: The Teleportation Circuit}
    
    In this section we show programs for the quantum teleportation circuit in each of the four languages for a side by side comparison. We remark that the QDK program shown is one of three programs needed to run the circuit, as discussed in the main body. The teleportation circuit is standard in quantum computing and sends an unknown state from one qubit|conventionally the first or top qubit in a circuit|to another|conventionally the last or bottom qubit in a circuit. Background information on this process can be found in any standard quantum computing or quantum mechanics resource. This quantum circuit is more involved than the very small programs shown in the main text and demonstrates some slightly more advanced features of each language|e.g., performing conditional operations. Note that the purpose of quantum teleportation is to transmit a qubit ``intact.'' We perform a measurement on the teleported qubit to verify the expected outcome is obtained.
    
    For completeness, we include a circuit diagram to make it clearer what the programs are doing. Unlike the main body of the text, this figure was made using the new \textsf{circuit\_drawer} released in Qiskit version 0.5.4. 
    
    \begin{figure}[h!]
    	\centering
        \includegraphics[scale=0.4]{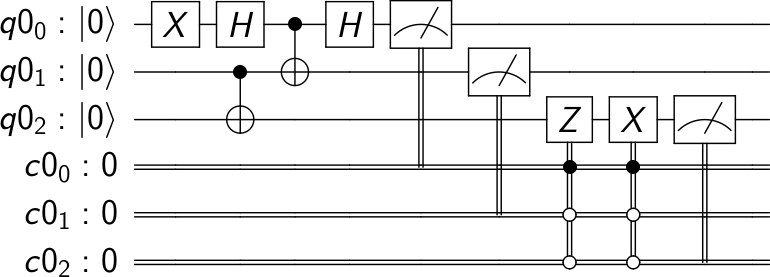}
        \caption{The teleportation circuit produced with the \textsf{circuit\_drawer} released in Qiskit v0.5.4.}
    \end{figure}

\nopagebreak

\begin{figure*}[h!]
	\centering
	\begin{tabular}{C{8cm}|C{8cm}}  
		\textbf{pyQuil} & \textbf{Qiskit} \\ \hline 
        {\footnotesize \begin{lstlisting}[language=Python]
#!/usr/bin/env python3
# -*- coding: utf-8 -*-

# ========================================
# teleport.py
#
# Teleportation circuit in pyQuil.
# ========================================

# ----------------------------------------
# imports
# ----------------------------------------

from pyquil.quil import Program
from pyquil import api
import pyquil.gates as gate

# ----------------------------------------
# program and simulator
# ----------------------------------------

qprog = Program()
qvm = api.QVMConnection()

# ----------------------------------------
# teleportation circuit
# ----------------------------------------

# perform X to teleport |1> to qubit three
qprog += gates.X(0)

# main circuit
qprog += [gates.H(1),
          gates.CNOT(1, 2),
          gates.CNOT(0, 1),
          gates.H(0),
          gates.MEASURE(0, 0),
          gates.MEASURE(1, 1)]

# conditional operations
qprog.if_then(0, gates.Z(2))
qprog.if_then(1, gates.X(2))

# measure qubit three
qprog.measure(2, 2)

# --------------------------------------
# run the circuit and print the results
# --------------------------------------

print(qvm.run(qprog))

# optionally print the quil code
print(qprog)\end{lstlisting}}
        &  {\footnotesize \begin{lstlisting}[language=Python]
#!/usr/bin/env python3
# -*- coding: utf-8 -*-

# ========================================
# teleport.py
#
# Teleportation circuit in Qiskit.
# ========================================

# ----------------------------------------
# imports
# ----------------------------------------

from qiskit import QuantumRegister, ClassicalRegister, QuantumCircuit, execute

# ----------------------------------------
# registers and quantum circuit
# ----------------------------------------

qreg = QuantumRegister(3)
creg = ClassicalRegister(3)
qcircuit = QuantumCircuit(qreg, creg)

# ----------------------------------------
# do the circuit
# ----------------------------------------

# perform X to teleport |1> to qubit three
qcircuit.x(qreg[0])

# main circuit
qcircuit.h(qreg[0])
qcircuit.cx(qreg[1], qreg[2])
qcircuit.cx(qreg[0], qreg[1])
qcircuit.h(qreg[0])
qcircuit.measure(qreg[0], creg[0])
qcircuit.measure(qreg[1], creg[1])

# conditional operations
qcircuit.z(qreg[2]).c_if(creg[0][0], 1)
qcircuit.x(qreg[2]).c_if(creg[1][0], 1)

# measure qubit three
qcircuit.measure(qreg[2], creg[2])

# -----------------------------------------
# run the circuit and print the results
# -----------------------------------------
result = execute(qcircuit, 'local_qasm_simulator').result()
counts = result.get_counts()

print(counts)

# optionally print the qasm code
print(qcircuit.qasm())

# optionally draw the circuit
from qiskit.tools.visualization import circuit_drawer
circuit_drawer(qcircuit)\end{lstlisting}
        }
	\end{tabular}
\end{figure*}


\newpage 

\begin{figure*}[h!]
	\centering
	\begin{tabular}{C{8cm}|C{8cm}}  
		\textbf{ProjectQ} & \textbf{Quantum Developer Kit} \\ \hline 
        {\footnotesize \begin{lstlisting}[language=Python]
#!/usr/bin/env python3
# -*- coding: utf-8 -*-

# ========================================
# teleport.py
#
# Teleportation circuit in ProjectQ.
# ========================================

# ----------------------------------------
# imports
# ----------------------------------------
from projectq import MainEngine
from projectq.meta import Control
import projectq.ops as ops

# ----------------------------------------
# engine and qubit register
# ----------------------------------------

# engine
eng = MainEngine()

# allocate qubit register
qbits = eng.allocate_qureg(3)

# ----------------------------------------
# teleportation circuit
# ----------------------------------------

# perform X to teleport |1> to qubit three
ops.X | qbits[0]

# main circuit
ops.H | qbits[1]
ops.CNOT | (qbits[1], qbits[2])
ops.CNOT | (qbits[0], qbits[1])
ops.H | qbits[0]
ops.Measure | (qbits[0], qbits[1])

# conditional operations
with Control(eng, qbits[1]):
    ops.X | qbits[2]
with Control(eng, qbits[1]):
    ops.Z | qbits[2]
    
# measure qubit three
ops.Measure | qbits[2]

# ----------------------------------------
# run the circuit and print the results
# ----------------------------------------
    
eng.flush()
print("Measured:", int(qbits[2]))\end{lstlisting}}
        &  {\footnotesize \begin{lstlisting}[language=C++]
// =========================================
// teleport.qs
//
// Teleportation circuit in QDK.
// =========================================

operation Teleport(msg : Qubit, there : Qubit) : () {
        body {

            using (register = Qubit[1]) {
                // get auxiliary qubit to prepare for teleportation
                let here = register[0];
            
                // main circuit
                H(here);
                CNOT(here, there);
                CNOT(msg, here);
                H(msg);

                // conditional operations
                if (M(msg) == One)  { Z(there); }
                if (M(here) == One) { X(there); }

                // reset the "here" qubit
                Reset(here);
            }

        }
    }
    
    operation TeleportClassicalMessage(message : Bool) : Bool {
        body {
            mutable measurement = false;

            using (register = Qubit[2]) {
                // two qubits
                let msg = register[0];
                let there = register[1];
                
                // encode message to send
                if (message) { X(msg); }
            
                // do the teleportation
                Teleport(msg, there);

                // check what message was sent
                if (M(there) == One) { set measurement = true; }

                // reset all qubits
                ResetAll(register);
            }

            return measurement;
        }
    }\end{lstlisting}
        }
	\end{tabular}
\end{figure*}

\end{document}